\documentclass[11pt]{article}
\usepackage{latexsym}
\usepackage{local}
\usepackage{term}
\usepackage{main}
\usepackage{prolog}

\begin{document}

\pagestyle{plain}     

\title{\bf A Logic-Based Approach to Data Integration}

\author{J.~Grant$^{3}$, \& J.~Minker$^{1,2}$\\
\small
{\em     jgrant@towson.edu,
         minker@cs.umd.edu} \\[3ex]
\large
$^1$%
Department of Computer Science \\
$^2$%
and Institute for Advanced Computer Studies \\
University of Maryland at College Park \\
College Park, Maryland, U.S.A. \\[1ex]
$^3$%
Department of Computer and Information Sciences and\\
Department of Mathematics\\
Towson University \\
Towson, Maryland, U.S.A. \\[1ex]
}

\date{}
\maketitle

\begin{abstract}

An important aspect of data integration involves answering queries
using various resources rather than by accessing database relations.
The process of transforming a query from the database relations to
the resources is often referred to as query folding or answering 
queries using views, where the views are the resources.
We present a uniform approach that includes as special cases much
of the previous work on this subject.
Our approach is logic-based using resolution.
We deal with integrity constraints, negation, and recursion also
within this framework.

\end{abstract}
 
\section{Introduction}\label{Sec:intro}
An important part of data integration involves answering queries
using various resources rather than by accessing database relations.
The process of transforming a query from the database relations to
the resources is often referred to as query folding or as answering queries
using views, where the views are the resources.
For instance, a database of interest to a user may be distributed over a 
network.  It is necessary to bring data distributed 
over a network to a user's machine so that the data may be
manipulated to answer user queries.  In a distributed 
environment it is likely that 
one will want to save answers to queries in 
the local user's machine so that if the same or a 
related query is posed to the distributed database, one can look
in the local machine's cached database for the answer, rather
than have to go out over the network to answer the query.
In this situation the resources are the cached relations and the
use of these resources is an important aspect of query optimization.
In some data integration systems the database relations are
themselves virtual and the data must be obtained from the resources.
Resources may also be materialized views.

Several researchers have considered various aspects of this problem.
In this paper we  present a logic-based approach to the query folding
problem using the method of resolution.  As a consequence, 
\begin{enumerate}
\item We obtain a uniform approach that includes as special cases much of the 
previous work on this topic.
\item If the algorithm finds a rewriting of the query, then we are
guaranteed that the answers are sound, that is, the answers are correct. 
\item The approach also allows us to determine under certain conditions
if the folded query contains all answers to the original query, that is, 
it is complete.
\end{enumerate}

We consider a database that consists of an extensional database ({\bf
EDB}), an intensional database ({\bf IDB}), a set of integrity
constraints ({\bf ICs}), and a set resources $({\bf Res_{DB}})$, where
the resources have been obtained by using resource rules.  These resources
are referred to as materialized views.  That is, they have been made
explicit in a local computer as a result, for example,  of an answer to
a conjunctive query.  The {\bf EDB}, {\bf IDB}, {\bf ICs} are part of a
conventional {\bf Datalog} database.

Section 2 describes related work, complexity results, systems and algorithms 
that have been developed with respect to the folding problem.
Section 3 provides background for the definitions and notations used in
the paper.
Section 4 contains several examples and our query folding algorithm.
The algorithm is logic-based and deals uniformly with integrity
constraints, extensional and intensional predicates and extends
the work in \cite{Q96}.
Functional and inclusion dependencies are considered in Section 5.
We show that our query folding algorithm handles all integrity
constraints in a uniform way without the need for specialized techniques
as in \cite{DGQ96,Gryz98} and \cite{DPT99:folding,PDST00:folding}, where the
latter also consider physical access structures, which we do not treat.
Section 6 deals with the case where resources are obtained by the
use of several definitions or queries.
Our work on multiple rules for the same resource
relates to the work in \cite{Duschka97,AGK99:disjviews,FG01:rewriting}.
Whereas they are concerned primarily with maximal containment, we are
concerned with a uniform method that checks both for soundness and completeness
of answers.  Section 7 discusses negation.  We show that the logic-based
framework handles stratified negation in both rules and intensional
predicates.  Our approach is different from the rewriting used in
\cite{FG01:rewriting}.  Recursion is considered in Section 8.  We differ from the work in
\cite{DG97:ans-queries-views} and \cite{DGL00:integration} by handling
a special case of recursive views as well as recursive queries.
We compare the contributions made in this paper with other efforts
in Section 8.  The paper is summarized in Section 9.

\section{Related Work}\label{Sec:related}
There has been a substantial amount of work done in connection with
data integration and query folding. 
Apparently the first papers to propose algorithms for query folding were
in \cite{LY85:folding,YL87}, where they developed an algebraic method. 
Other early work was by \cite{TSI94:folding} and by \cite{CKPS95}.
An algorithm for rewriting conjunctive
queries over non-recursive databases was provided in
\cite{Q96}.  In \cite{DGQ96,Gryz98} it is shown how to use
materialized views in the presence of functional and inclusion dependencies.
Additional work on conjunctive query optimization 
and on information integration appears in
\cite{Ullm97,LRO96a,LRO96b,DG97:ans-queries-views}.
Levy et al.
\cite{LMSS95:views} showed that the question of determining whether a
conjunctive query can be rewritten to an equivalent conjunctive query
that only uses views is NP-complete.  This work was extended in
\cite{RSU95} to include binding patterns in view definitions.  
Duschka \cite{Duschka97:AAAI}, discusses the concept of 
{\em local completeness}, where it is known that some subset of the 
data an information source stores is complete, although the entire data 
stored by the information source might not be complete.
\cite{DG97:ans-queries-views} were the first to extend the work to general recursive queries.
They show that the problem of whether a Datalog program can be rewritten into an equivalent
program that only uses views, is undecidable. \cite{DGL00:integration} 
also discuss
answering queries using views to recursive queries for
\datalog\ programs.  Duschka and Levy \cite{DL97} introduce the new class of 
{\em recursive} query plans for information gathering.  Plans are 
extended to be recursive sets of function-free Horn clauses.  
In his thesis, Duschka \cite{Duschka97} deals with
multiple definitions of the same resource and shows how to obtain
a maximally contained query using these definitions.  Duschka and
Genesereth \cite{DG98:disj} were the first to publish results concerning
how to handle multiple definitions of the same resource predicate.
Afrati et al.
\cite{AGK99:disjviews} extend this work to disjunctive queries and
related results.  Popa and his co-authors, \cite{DPT99:folding,PDST00:folding},
showed how to do query folding with some forms of integrity constraints 
using the so-called `chase method'.
Flesca and Greco \cite{FG01:rewriting} deal with
disjunction and negation and formulate their answers in terms of
classical and default negation.
The complexity of answering queries using
materialized views for conjunctive queries with inequality, positive
queries, \datalog\, and first-order logic is addressed in
\cite{AD98:materialized-views}.  The paper \cite{Levy01:survey} surveys 
the methods proposed for answering queries using views.  
See Ullman, \cite{Ullm97}, for a survey of work concerning 
information-integration tools
to answer queries using views that represent the capabilities of information
resources.  The formal basis of techniques related to containment
algorithms for conjunctive queries and/or Datalog programs is discussed there.  
Approaches taken by AT\&T Labs's {\em Information Manifold} and the Stanford
{\em Tsimmis}, \cite{GM*95:Tsimmis} project are compared.
Levy in \cite{levy-ay:2000a} describes several algorithms proposed for data
integration: the bucket and inverse-rules algorithms. 

Several papers address complexity problems associated with folding,
Chandra and Merlin, \cite{CM77} showed that the query containment 
problem is NP-complete.  Several subclasses of conjunctive queries 
were identified  that have polynomial-time containment algorithms
\cite{ASU79,ASU79a,JK83}.

The query folding problem is thus at least NP-hard.  It has been
shown to be NP-complete for conjunctive queries and resources 
in \cite{LMSS95:views}.
Many variants of the problem of answering queries using views are
discussed in \cite{Levy01:survey}.  The problem was shown to be NP-complete
even when queries describing the sources and the user query are
conjunctive and do not contain interpreted predicates (\cite{LMSS95:views}).
\cite{LMSS95:views} further show that in the case of conjunctive queries, the
candidate rewritings can be limited to those that have at most
the number of subgoals in the query.  The complexity of the problem is
polynomial in the number of views (i.e., the number of data sources in
the context  of data integration).
Since query containment is a special case of query folding,
Qian's algorithm, \cite{Q96} degenerates to a polynomial-time containment 
algorithm for the class of acyclic conjunctive queries.

Abiteboul and Duschka \cite{AD98:materialized-views}
show that recursion and negation in the view definition
lead to undecidability.  They show that the {\em Closed World 
Assumption (CWA)} complicates the problem.
Under the {\em Open World Assumption (OWA)}, the certain answers 
in the conjunctive view definitions/Datalog queries case can be 
computed in polynomial time.  On the other hand, the conjunctive 
view definitions/conjunctive queries case is co-NP-complete under 
the CWA.  They prove that inequalities (a weak form of negation) 
lead to intractability.  Even under the OWA, adding inequalities 
to the queries, or disjunction to the view definitions make the 
problem co-NP hard.  In his thesis, Duschka, \cite{Duschka97}
provides a summary of results in complexity associated with these
results.

Chekuri and Rajaraman, \cite{CR97:contain},
present polynomial-time algorithms to test the containment of
an arbitrary conjunctive query in an acyclic query and to minimize
an acyclic query.  They generalize the query containment and
minimization algorithms to arbitrary queries.
They consider the problem of finding an equivalent rewriting of a
conjunctive query $Q$ using a set of views $\mathcal V$ defined by
conjunctive queries, when $Q$ does not have repeated predicates, and
show how their algorithms for query containment can be modified for
this problem.  A restricted variant of this problem, where neither
$Q$ nor the views in $\mathcal V$ use repeated predicates, is known to
be $NP-complete$ \cite{LMSS95:views}.

Duschka and Genesereth, \cite{DG98:disj},
treat views that may be defined by disjunction.
Their focus is on maximal query containment. They show a duality 
between a query plan being maximally contained in a query and this
plan computing exactly the certain answers.  
They show that the plan they generate is maximally contained in 
the query and that the disjunctive plan can be evaluated in co-NP time.
The complexity results described above also apply to the problems that we
discuss in this paper.

Several systems, and a number of algorithms have been implemented
for the folding problem.
Levy, Rajaraman, and Ordille \cite{LRO96b}, developed the
{\em Information Manifold System} at AT\&T Labs. The system incorporates 
the {\em bucket algorithm}, which controls search by first considering 
each subgoal in a query in isolation, and creating a bucket that 
contains only the views relevant to that subgoal.  The algorithm 
then creates rewritings by combining one view from each bucket.

Qian and Duschka and Genesereth \cite{Q96,DG97:ans-queries-views,DG97:infomaster,Duschka97}, 
are responsible for the {\em InfoMaster System}. 
They use the {\em inverse-rules algorithm}, and consider rewritings
for each database relation independent of any particular query.  Given a
user query, these rewritings are combined appropriately.  They show that
rewritings produced by the inverse-rules algorithm need to be further
processed in order to be appropriate for query evaluation.  They show this
additional processing step duplicates much of the work done in the second phase
of the bucket algorithm.  The bucket algorithm is also shown to have
several deficiencies and does not scale up.  Details of these algorithms are
presented by Levy in \cite{levy-ay:2000a}. 

Pottinger and Levy \cite{PL00:MiniCon} have developed a scalable algorithm for 
answering queries with views.  They describe and analyze the bucket and 
the inverse-rule algorithms.  They then describe the MiniCon algorithm, 
for finding the maximally contained rewriting of a conjunctive query 
using conjunctive views.  They provide the first experimental study of 
algorithms for answering queries. They show that the MiniCon algorithm 
both scales up and significantly outperforms the previous algorithms.  
They further develop an extension of the MiniCon algorithm to handle 
comparison predicates, and show its performance experimentally.

Afrati, Li, and Ullman \cite{ALU01:CoreCover}, discuss generating efficient, equivalent
rewritings using views to compute the answer to a query.  Each rewriting of a
query is passed as a logical plan to an {\em optimizer}, which translates
the rewriting to a {\em physical plan}.  Each physical plan accesses the 
stored ("materialized") views, and applies a sequence of  relational 
operators to compute the answer to the original query.  They consider three
cost models for evaluating the efficacy of a plan.  They develop and 
experimentally evaluate an efficient algorithm, {\em CoreCover}, based on
a simple cost model $M_1$ that counts only the number of subgoals in
a physical plan.

\section{Background}\label{Sec:back}
This section contains a summary of the background and notation
used in this paper.
We use the language and terminology of logic databases%
\index{database!logic}
(also known as deductive databases)
(\cite{Das92:deductive},
\cite{Llo87:foundations},
and
\cite{LMR92:book}).
Logic databases%
\index{database!logic}
express data, rules%
\index{rule}
(views)%
\index{view},
integrity constraints%
\index{integrity!constraints}
and queries in first-order logic%
\index{logic!first-order}
(\cite{BJ89:computability}
and
\cite{Llo87:foundations}).
\footnote{%
Some proposals for integrity constraints express them in a 
higher-order logic,%
\index{logic!higher order}
while keeping the database---%
the facts, rules, and, sometimes, queries---%
in first-order.
}

We use standard syntax for first-order logic,%
\index{logic!first-order}
with the usual symbols for variables, connectives, quantifiers,
punctuation, equality, constants, and predicates.
The notions of term, formula, and sentence
(a formula with no free variables)
are defined in the usual way.
We do not necessarily restrict formulas to be function-free.%
\index{formula!function-free}
We shall make it clear when we utilize function symbols.
Any formula may be considered as a {\em query}%
\index{query}
\index{formula!query|see{query}}
(although, below, we restrict the form of standard queries).
A formula is called {\em ground}%
\index{formula!ground}
if it contains no variables.

A {\em substitution}%
\index{substitution}
is a set of {\em substitution pairs},
for example, $\{\binding{X}{a}, \binding{Y}{b}\}$,
such that every element of a substitution pair is a variable or constant
(or, more generally, a term),
and such that the collection of left-hand sides of the substitution pairs---%
\var{X}\ and \var{Y}\ in our example---%
are unique variables.
A substitution {\em applied} to a formula is a rewrite of the formula 
by replacing any occurrence in the formula of a left-hand element
from the substitution by its right-hand counterpart, in parallel.
Let the formula \set{F}\ be \atom{p}{X,Y}\ and the substitution $\theta$
be again $\{\binding{X}{a}, \binding{Y}{b}\}$.
The substitution $\theta$ applied to formula \set{F},
written as $\set{F}\theta$,
is the formula \atom{p}{a,b}.

We assume that the reader is familiar with the unification algorithm \cite{Robi65}.
The unification algorithm takes a set of relations with the same relation name and
attempts to find a substitution for the variables that will make the relations all
identical.


An important class of sentence is the {\em clause}.%
\index{clause}
A clause has the general form:
\begin{equation}
\forall.
A_1 \vee \ldots \vee A_k \vee
\neg A_{k+1} \vee \ldots \vee \neg A_n
\end{equation}
in which each $A_i$ is an atomic formula%
\index{formula!atomic},
for \range{i}{1}{n},
and in which the variables are understood to be universally quantified
(denoted by `$\forall$').
Any clause%
\index{clause}
can be written in a logically equivalent form as an implication.
\begin{equation}
\forall.
A_1 \vee \ldots \vee A_k 
\leftarrow A_{k+1} \wedge \ldots \wedge A_n.
\end{equation}
This is often written in further shorthand as
\begin{equation}
A_1,\ldots,A_k
\leftarrow A_{k+1},\ldots,A_n.
\end{equation}
in which disjunction is assumed on the left-hand side
of the implication arrow,
conjunction on the right-hand side,
and the universal quantification is understood.
A clause%
\index{clause}
in this form is also called a {\em rule}.%
\index{rule}
The collection of atoms on the left-hand side
($\var{A}[1],\ldots,\var{A}[k]$)
is called the {\em head}%
\index{rule!head of}
of the rule,
and the collection of atoms on the right-hand side
($\var{A}[k+1],\ldots,\var{A}[n]$)
the {\em body}.
When $k=n$, the body is empty,
and when $k=0$, the head is empty.
A {\em Horn} rule%
\index{rule!Horn}
has at most one atom in the head: $k \le 1$.
A {\em definite} clause%
\index{clause!definite}
has exactly one atom in the head: $k=1$.
A {\em ground} rule contains no variables. 
In a rule a variable is called {\em limited} if it appears in the body of an
ordinary (non built-in) atom or in an equality with a constant
or a variable that is limited.
A rule is called {\em safe} if all its variables are limited.

A rule with an empty head%
\index{rule!head of!empty}
is generally considered to be a {\em query} or an integrity constraint.
\index{query}
An {\em answer} to a query%
\index{query!answer to}
is a ground substitution%
\index{substitution!ground}
of the query%
\index{query}
formula
such that the resulting ground formula%
\index{formula!ground}
is {\em true} with respect to the database;
that is, the grounded query formula%
\index{formula!ground} is {\em logically entailed} by the database.
A ground rule with an empty body%
\index{rule!body of!empty}
is called a {\em fact}.
Definite rules%
\index{rule!definite}
have a clear procedural interpretation.
Consider
\begin{equation}
A \leftarrow B_1,\ldots,B_n.
\end{equation}
We call this clause a rule {\em for} \var{A}.
The above rule can be interpreted to say that \var{A}\ is shown
(or {\em proven}) whenever all the \var{B}[i]'s are shown ({\em proven}).%
\footnote{%
The interpretation for disjunctive rules%
\index{rule!disjunctive} 
is less apparent.
Essentially, a disjunctive rule states that at least {\em one}
of the atoms in the rule's head is {\em proven}
whenever all the atoms in the body are.
}
Rules are essentially {\em views},%
\index{view}
in the parlance of relational databases.%
\index{database!relational}
Logically,
rules are more expressive than views in relational databases 
because recursion%
\index{recursion}
is permitted.%
\footnote{%
The \sql-3 standard extends \sql\ to support recursion
\cite{MS93:SQL}, however.
So once \sql-3 becomes the standard,
this difference in expressiveness will go away,
since any relational database%
\index{database!relational} 
that supports \sql-3 will,
in fact, be a deductive database%
\index{database!deductive}
system.
}
A fact then,
having no conditions in the body of its ``rule'',
is simply interpreted as {\em true}.
%
We define an {\em expanded rule}\label{expanded} (\cite{Cha85:phd,Chak88}) to be one in which 
all predicates have been expanded.  An {\em expanded predicate} is one in which
all constants and repeated variables have been replaced
by unique new variables, and the appropriate equalities have been added to the
body of the rule in which the predicate appears (\cite{CGM90}).  This is 
related to the term {\em rectified} set of rules where the head of each rule
in the set is identical with each argument a distinct variable (\cite{Ull88:principles}).

A database may then be defined as a collection of rules%
\index{rule}
and facts.
When all the rules%
\index{rule!definite}
and facts are definite
 (that is, the rules and facts have at most a single atom in the heads
of the clauses that define them),
the database is called {\em definite}.%
\index{database!definite}
It is called {\em disjunctive}%
\index{database!disjunctive}
(or {\em indefinite}) otherwise.
We call the language in which the database is written with definite clauses
\index{clause}
as defined above
\datalog%
\index{Datalog}
\cite{Ull88:principles}.  Recall that terms in clauses are function-free,
as noted above, hence all \datalog\ terms are function-free.
When disjunctive clauses are permitted for rules or facts (and whose terms are 
function-free), we call the 
language Disjunctive \datalog\ \cite{EitGotMan97,LMR92:book}.
%
A database \DB\ often is defined as consisting of two parts:
\begin{itemize}
\item the extensional database,%
	\index{database!extensional ({\bf EDB})} 
	{\bf EDB},
	and
\item the intensional database,%
	\index{database!intensional ({\bf IDB})}
	{\bf IDB}.
\end{itemize}
The {\bf EDB} is the database's collection of {\em facts}.
The {\bf IDB} is the database's collection of rules.
(We soon redefine databases to have two additional components,
the set of the database's integrity constraints ({\bf ICs}) and the set of 
resource rules (${\bf Res_{DB}}$)).

Conventionally,
negative data is not represented explicitly in a logic database.%
\index{database!logic}
There are several standard approaches to allow negative data to be inferred.
The {\em closed world assumption}%
\index{closed world assumption (\cwa)}
(\cwa)
is a default rule for the inference of negative facts
\cite{R78}.
For any ground atom \var{A},
the negation%
\index{negation|(}
of \var{A}\ is accepted as {\em true}
if \var{A}\ is not provable from the database.
The set of all negated atoms inferable in this way
is written as \op{CWA}{\DB}.
Another approach to negation
is the {\em Clark completion}%
\index{completion}
of a database \cite{C78}.
This formalizes the concept that the set of tuples {\em true}
for a predicate is precisely the set that can be proven to be {\em true}
via the facts and rules.
In brief,
this is accomplished by adding a formula to the database
for each predicate
(to correspond with the collection of rules%
\index{rule}
for that predicate),
to supply the logical {\em only if} half of the definition
of the predicate.
Certain negated facts are then {\em deducible}
from the {\em completed} database,%
\index{completion}
the database with these ``only if'' formulas added.  
We refer to \cite{C78} for the precise definition.
In our application we will typically have a situation where a
resource predicate is defined by some extensional predicates, such as\\
\begin{equation}\label{skolem}
r(X,Y) \leftarrow h(X,Z), k(Z,Y)
\end{equation}
where $r$ is the resource predicate and $h$ and $k$ are extensional
predicates, saying essentially that if $<a,b>$ is in the join of $h$ and
$k$, it is in $r$. The Clark completion changes the implication to
an equivalence to
say that 
$<a,b>$ is in the join of $h$ and $k$ if and only
if $<a,b>$ is in $r$.
When we compute the Clark completion of a rule such as in clause~\ref{skolem}
we obtain two rules, one for $h$ and one for $k$.  The variable $Z$ in 
clause~\ref{skolem} represents an existensionally quantified variable whose
value depends on the variables $X$ and $Y$.  When we obtain the only-if part
of the Clark completion, namely, the rules with the implication arrow reversed,
the predicates with the variable $Z$ appear in the heads of the clauses and are
replaced by the Skolem function $f(X,Y)$.  Thus, the only-if portion of the
Clark completion become:
\begin{equation}
h(X,f(X,Y)) \leftarrow r(X,Y)
\end{equation}
\begin{equation}
k(f(X,Y),Y) \leftarrow r(X,Y)
\end{equation}
In the text, when there is no loss of information, we omit the variable
portion of the Skolem functions.  For example, we replace $f(X,Y)$ by $f$.

%
%
%

In our formulas we allow built-in predicates, such as $=, <, >$.  When built-in
predicates occur in formulas the appropriate axioms need to be added to the
theory.  The axioms for equality, for example, are given below.

{\bf Equality Axioms:}\\
\noindent
\hspace*{0.5in} $\forall X (X = X)$\\
\hspace*{0.5in} $\forall X \forall Y ((X=Y) \rightarrow~(Y = X))$\\
\hspace*{0.5in} $\forall X \forall Y \forall Z ((X=Y) \wedge (Y=Z)
\rightarrow (X=Z))$\\
\hspace*{0.5in} $\forall X_{1} \cdots \forall X_{n} (P(X_{1},
\cdots, X_{n}) \wedge$
$(X_{1} = Y_{1}) \wedge \cdots \wedge (X_{n} =
Y_{n} )  \rightarrow P (Y_{1}, \cdots Y_{n} ))$\\
We will discuss the use of equality axioms when they are needed in
proofs.  

So far,
we have assumed that the body of a database rule (clause)%
\index{clause}
contains only positive atoms.
However,
it is useful sometimes to define database rules
that allow negated atoms in the body of a rule.
We need default negation in logic databases%
\index{database!logic}
if we want to subsume
the relational algebra, which includes set difference.
We can extend deductive databases%
\index{database!deductive}
with default negation.
A rule which has a negated atom, i.e. an atom preceded by 
{\em not} in its body
is called a {\em normal} rule,%
\index{rule!normal}
and deductive databases that have normal rules
are called {\em normal} databases.%
\index{database!normal}
We call \datalog%
\index{Datalog}
that has been extended with default negation \datalogN.
For example, the normal rule
\begin{equation}
p(X) \leftarrow not~q(X).
\end{equation}
is interpreted, in general, to mean that,
for any constant \var{a},
if \atom{q}{a}\ is {\em not true}
(or cannot be {\em proven} to be {\em true}),
then \atom{p}{a}\ is {\em true}.

We write this negation with {\em not} rather than with
the symbol for logical negation,%
\index{negation!logical}
`$\neg$',
and refer to it as {\em default negation}.%
\index{negation!default}
This is because most semantics%
\index{semantics}
that have been defined
for normal databases,%
\index{database!normal}
interpret the use of default negation
differently from one another and from logical negation.
There are a number of semantics that have been defined
for normal databases,
and no one semantics%
\index{semantics}
is universally accepted.
Also, since the notion of default negation is generally based on provability,
not logical truth, such default negation is beyond first-order logic.%
\index{logic!first-order}%
\footnote{%
We still speak in terms of first-order%
\index{logic!first-order}
logic even for normal 
databases,%
\index{database!normal}
as most of the first-order framework of deductive databases%
\index{database!deductive}
remains applicable.
}

Thus it is not equivalent to exchange the rules in the {\bf IDB}
that use default negations
with seemingly equivalent disjunctive rules,
in which the negated atoms in the body have been moved to the head.
Consider the following example.\\
{\bf $DB_1$} contains  two clauses: (1) $p \leftarrow q, not~r$, and  (2) $q$.\\
{\bf $DB_2$} contains  two clauses: (1) $p \vee r \leftarrow q$, and  (2) $q$.\\
That is, 
let \DB[1]\ consist of the single rule for the predicate $p$\
and the single fact for the predicate $q$,
and \DB[2]\ consist of the single disjunctive rule%
\index{rule!disjunctive}
and the single fact for the predicate $q$.
If the rule in \DB[1]\ were written with logical negation
instead of default negation,
\DB[1]\ and \DB[2]\ would be logically equivalent.
However, in \DB[1],
we should be able to infer $p$,
because the fact $q$ can be inferred
(it is a fact),
and the fact $r$ {\em cannot} be inferred,
(thus, $not~r$ can be assumed {\em true} by default).
In \DB[2],
$p$ cannot be inferred.
Only the weaker, disjunctive fact
$p \vee r$ 
can be inferred.
Note that default negation results in non-monotonicity.%
\index{logic!non-monotonic}
If we were to {\em add} the fact $r$ to \DB[1]\ above,
we would no longer be able to infer $p$.


The intuition behind the use of default negation
becomes confused when it is combined with recursion.%
\index{recursion}
One solution to this confusion is simply not to allow
recursive definitions through negation.
A canonical example of recursion through negation is

\begin{equation}
p(X) \leftarrow not~q(X). 
\end{equation}
\begin{equation}
q(X) \leftarrow not~p(X). 
\end{equation}

The restriction not to allow recursion%
\index{recursion}
through negation leads to what are called {\em stratified} databases,%
\index{database!stratified}
and such databases have a unique standard model
called the {\em perfect model}%
\index{model!perfect}
of the database.
(Stratified databases are defined in \cite{ABW88,LMR92:book}).
In some cases,
a non-stratified database may also have a unique standard model.%
\index{model!unique}
Some of these cases may be captured by the concept of {\em stable} database.%
\index{database!stable}
Two important model semantics for normal databases,%
\index{database!normal}
and normal logic programs,
are the {\em well-founded semantics}%
\index{semantics!well-founded}
\cite{VRS91}
and the {\em stable model semantics}%
\index{semantics!model!stable}
or the semantically equivalent {\em well-supported model semantics}%
\index{semantics!model!well-supported}
\cite{Fag91:fixpoint,GL88:stable}.
\cite{Mink96:retrospective} provides a retrospective on work in semantics
for logic programs and deductive databases.%
\index{database!deductive}


\section{Query Folding}\label{Sec:fold}
This section contains the basic material on query folding. 
We assume that the resource rules define the predicates of the data
sources that are conveniently available while the {\bf EDB} and {\bf IDB}
predicates may take longer to use or may be unavailable.
Consequently, a query is optimized in the sense that it has been rewritten 
using the data sources, which presumably are readily available.  The
folded query can then be optimized by other well-known techniques.
In this section we provide an algorithm for such query rewriting in a
special case.
In later sections we extend this algorithm to more complex databases.
We start by giving the restrictions on the type of database we consider
in Sections 4 and 5.
Our query folding algorithm is illustrated on an example before it is described.
We end this section by giving several additional examples.

\subsection{Database Restrictions}\label{restrict}

From now on a database will consist of four parts: 
the {\bf EDB}, {\bf IDB}, {\bf IC}, and ${\bf Res_{DB}}$.
We assume that the resource rules define the predicates of the data
sources that are conveniently available while the {\bf EDB} and {\bf IDB}
predicates may take longer to use.
In this section we provide an algorithm for such query rewriting in a
special case.
In later sections we will extend this algorithm to more complex databases.

We place the following conditions on the database in this and the
following section.
\begin{enumerate}
  \item No formula contains negation.
  \item Each {\bf IDB} predicate may be defined by multiple safe,
conjunctive, non-recursive function-free Horn rules.
  \item Each distinct ${\bf Res_{DB}}$ predicate is defined by a single 
safe conjunctive function-free Horn formula on {\bf EDB} and/or {\bf IDB} predicates.
  \item Each {\bf IC} clause is a safe function-free Horn formula of the
form $~~~G \leftarrow F~~~$
where $G$ is either empty or has one {\bf EDB} predicate 
and $F$ is a conjunction of {\bf EDB} predicates.
  \item Each query has the form $~~~q: \leftarrow G~~~$ where $G$ is a
conjunction of {\bf EDB} and {\bf IDB} predicates.
  \item The database includes axioms for built-in predicates as needed. 
\end{enumerate}

When the {\bf IDB} is non-recursive, it has been shown \cite{Reit78A} that
the rules can be compiled so that every {\bf IDB} predicate can be written
as a set of rules, each rule in terms only of {\bf EDB} predicates.  We 
assume in this section that the compiled rules replace the original rules.
Hence, deduction using {\bf IDB} predicates is effectively one-step.

In the following we consider the concept of {\em bounded recursion}.
Minker\index{Minker, J.} and
Nicolas\index{Nicolas, J.-M.} \cite{Mink82A} were the first to show
that there are forms of
rules that lead to {\em bounded recursion}\index{bounded recursion,
definition}.  That is, the deduction process using these rules must
terminate in a finite number of steps.  This work has been extended by
Naughton\index{Naughton, J.F.} and Sagiv\index{Sagiv, Y.}
\cite{Naug87}.  
We illustrate here one special case of bounded recursion, namely,
singular rules.

A recursive rule is
{\em singular}\index{singular rule, definition} if it is of the form\\
\begin{center}

$R \leftarrow F \wedge R_1 \wedge \wedge \ldots \wedge R_n$,

\end{center}

\noindent
where $F$ is a conjunction of possibly empty base (i.e. {\em EDB}) relations
and $R, R_1, R_2, \ldots, R_n$ are atoms that have the same relation name iff:
\begin{enumerate}

\item each variable that occurs in an atom $R_i$ and does not occur in $R$
only occurs in $R_i$;

\item each variable in $R$ occurs in the same argument position in any
atom $R_i$ where it appears, except perhaps in at most one atom $R_1$ that
contains all of the variables of $R$.

\end{enumerate}

\noindent
Thus, the rule
\begin{center}

$R(X,Y,Z) \leftarrow R(X, Y',Z), R(X,Y,Z')$

\end{center}
\noindent
representing the multivalued dependency, $R:X\rightarrow\rightarrow Y$ is 
singular since (a) $Y'$ and $Z'$ appear respectively in the first and
second atoms in the head of the rule (condition 1), and (b) the variables
$X,Y,Z$ always appear in the same argument position (condition 2).
It is known that all singular rules have bounded recursion.

We now specify three cases for the {\bf IC} since each case  has to
be handled in a different manner in the folding algorithm:\\
\newpage
{\bf Cases of Integrity Constraints (ICs)}\label{iccases}
\begin{enumerate}
\item Case 1: The {\bf ICs} have no recursion and no built-in predicate
in the head of a clause.  

\item Case 2: The {\bf ICs} have bounded recursion and no built-in predicate in
the head of a clause.  For example this is the case if there is a multivalued
dependency.

\item Case 3: Either the {\bf ICs} are recursive, or there is a 
built-in predicate in the head of a clause.
For example, for a functional dependency, an induced recursion
may arise since the equality axioms are recursive.
\end{enumerate}

\subsection{Illustrative Example}

We start by illustrating our algorithm on a simple example.
This example has a simple integrity constraint.
We deal with functional and inclusion dependencies in the next section.
We write the formal description afterwards and show how it subsumes
other algorithms used for this type of query rewriting.
\begin{example}\label{simple}
{\bf EDB}: $p_1(X,Y,Z)$, $p_2(X,U)$, $p_3(X,Y)$\\
{\bf IDB}: $\emptyset$\\
{\bf IC}: $p_3(X,Y) \leftarrow p_1(X,Y,Z), Z > 0$\\
${\bf Res_{DB}}$:  $r(X,Y,Z) \leftarrow p_1(X,Y,Z), p_2(X,U)$\\
{\bf Query}: $q(X,Y): \leftarrow p_1(X,Y,Z), p_2(X,U), p_3(X,Y), Z > 1$\\
\\
The first step involves reversing the resource rules to define the
{\bf EDB} predicates in terms of the resource predicates.
This process supplies the only if half of the definition of the
resource predicates; so we call these rules the Clark Completion
resource rules.
In this example we obtain for
the Clark Completion resource rules:\\\\
$CCrr1: p_1(X,Y,Z) \leftarrow r(X,Y,Z)$\\
$CCrr2: p_2(X,f(X,Y,Z)) \leftarrow r(X,Y,Z)$\\

\begin{figure}
\label{lit-elim}
\setlength{\unitlength}{0.00083300in}%
\begingroup\makeatletter\ifx\SetFigFont\undefined
\def\x#1#2#3#4#5#6#7\relax{\def\x{#1#2#3#4#5#6}}%
\expandafter\x\fmtname xxxxxx\relax \def\y{splain}%
\ifx\x\y   
\gdef\SetFigFont#1#2#3{%
  \ifnum #1<17\tiny\else \ifnum #1<20\small\else
  \ifnum #1<24\normalsize\else \ifnum #1<29\large\else
  \ifnum #1<34\Large\else \ifnum #1<41\LARGE\else
     \huge\fi\fi\fi\fi\fi\fi
  \csname #3\endcsname}%
\else
\gdef\SetFigFont#1#2#3{\begingroup
  \count@#1\relax \ifnum 25<\count@\count@25\fi
  \def\x{\endgroup\@setsize\SetFigFont{#2pt}}%
  \expandafter\x
    \csname \romannumeral\the\count@ pt\expandafter\endcsname
    \csname @\romannumeral\the\count@ pt\endcsname
  \csname #3\endcsname}%
\fi
\fi\endgroup
\begin{picture}(2944,5442)(1126,-5197)
\thicklines
\put(2851, 14){\line( 0,-1){750}}
\put(2851,-736){\line( 4, 1){1129.412}}
\put(2851,-1036){\line( 0,-1){750}}
\put(2851,-2086){\line( 0,-1){750}}
\put(2933,-2789){\line( 4, 1){1129.412}}
\put(2851,-3136){\line( 0,-1){750}}
\put(2851,-3886){\line( 2, 1){1110}}
\put(2851,-4186){\line( 0,-1){750}}
\put(1126, 89){\makebox(0,0)[lb]{\smash{\SetFigFont{12}{14.4}{rm}$Query: \leftarrow p_1(X,Y,Z), p_2(X,U), p_3(X,Y) ,Z>1$}}}
\put(3976,-361){\makebox(0,0)[lb]{\smash{\SetFigFont{12}{14.4}{rm}$IC: p_3(X,Y) \leftarrow p_1(X,Y,Z), Z>0$}}}
\put(1576,-961){\makebox(0,0)[lb]{\smash{\SetFigFont{12}{14.4}{rm}$ \leftarrow p_1(X,Y,Z), p_2(X,U), p_1(X,Y,Z), Z>0, Z>1$}}}
\put(1576,-2011){\makebox(0,0)[lb]{\smash{\SetFigFont{12}{14.4}{rm}$ \leftarrow p_1(X,Y,Z), p_2(X,U),  Z>1$}}}
\put(1651,-3061){\makebox(0,0)[lb]{\smash{\SetFigFont{12}{14.4}{rm}$ \leftarrow r(X,Y,Z), p_2(X,U),  Z>1$}}}
\put(4051,-2386){\makebox(0,0)[lb]{\smash{\SetFigFont{12}{14.4}{rm}$CCrr1: p_1(X,Y,Z) \leftarrow r(X,Y,Z)$}}}
\put(1651,-4111){\makebox(0,0)[lb]{\smash{\SetFigFont{12}{14.4}{rm}$ \leftarrow r(X,Y,Z), r(X,Y,Z),  Z>1$}}}
\put(3976,-3286){\makebox(0,0)[lb]{\smash{\SetFigFont{12}{14.4}{rm}$CCrr2: p_2(X,f(X,Y,Z)) \leftarrow r(X,Y,Z)$}}}
\put(3676,-3661){\makebox(0,0)[lb]{\smash{\SetFigFont{12}{14.4}{rm}$\{U/f(X,Y,Z)\}$}}}
\put(1651,-5161){\makebox(0,0)[lb]{\smash{\SetFigFont{12}{14.4}{rm}$ \leftarrow r(X,Y,Z),  Z>1$}}}
\put(3151,-4561){\makebox(0,0)[lb]{\smash{\SetFigFont{12}{14.4}{rm}factor}}}
\put(3151,-1411){\makebox(0,0)[lb]{\smash{\SetFigFont{12}{14.4}{rm}factor and subsume}}}
\end{picture}
\caption{Literal Elimination Example}
\end{figure}
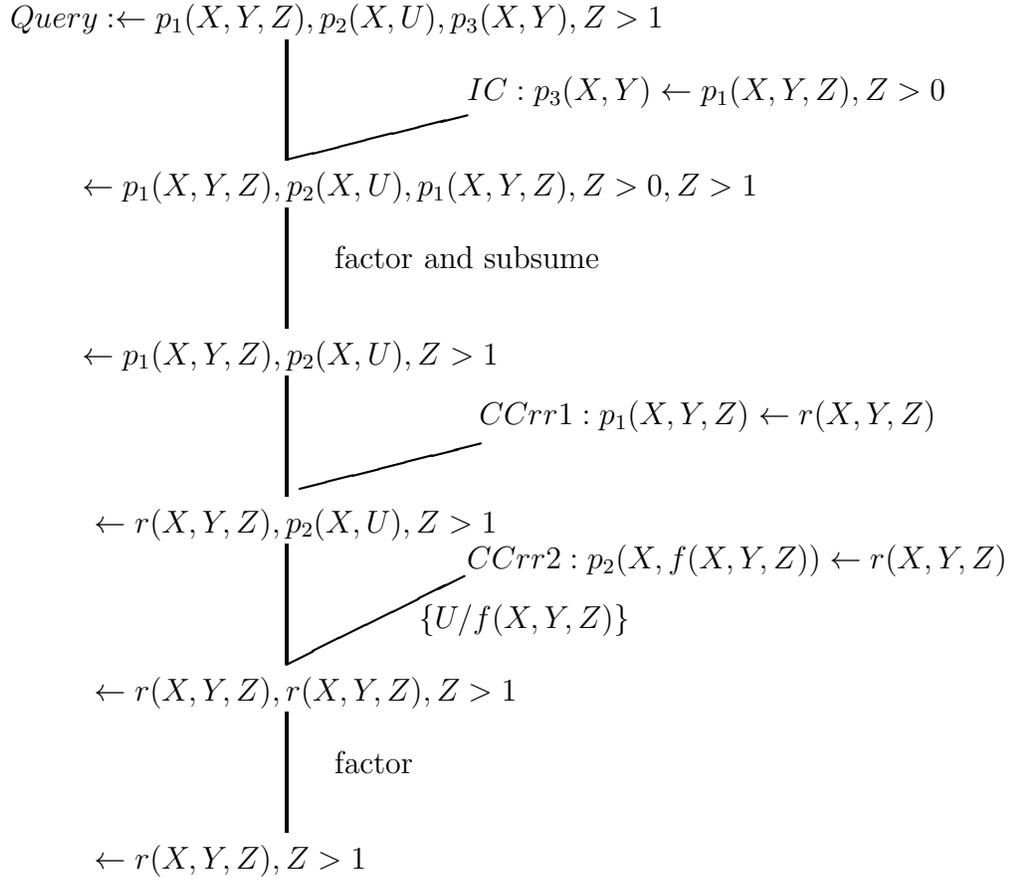

Figure~\ref{lit-elim} 
shows the derivation starting with the query 
as the top clause of a linear resolution tree.
Both the {\bf IC} and the Clark Completion resource
rules are used.
The clause $~Z > 0~$ is subsumed by the clause $~Z > 1~$.
The final rewritten query at the bottom of the tree\\
$\leftarrow r(X,Y,Z), Z > 1$,\\
contains only the $r$ predicate and the evaluable predicate $>$.
Using this query we obtain correct answers to the original query.

This example uses an integrity constraint to obtain a
clause that contains only a resource predicate and an evaluable predicate.
When the theory is Horn, Reiter has shown \cite{Reit78A} that the use
of integrity constraints is not necessary to obtain answers.  In
this case, if the integrity constraint were not used, the clause
at the end of the proof tree would have been a partial folding consisting
of a resource predicate, an {\bf EDB} predicate and an evaluable predicate.
Using the integrity constraint eliminates the {\bf EDB} predicate and
provides an optimization step, as in the case of semantic query optimization
\cite{CGM90}.
We can illustrate another aspect of semantic query optimization by changing
the integrity constraint in this example to:\\\\
{\bf IC}: $ \leftarrow p_2(X,U), p_3(X,Y)$\\\\
This integrity constraint subsumes the query; hence the query has no
answers and there is no need to try to fold the query.
\end{example}

\subsection{Query Folding Algorithm}


At this point we describe the first version of the query folding algorithm,
where the database satisfies the six conditions given at the beginning of
Section~\ref{restrict}.  In particular each resource predicate is 
defined by a single safe conjunctive formula on {\bf EDB}.

As mentioned in Example~\ref{simple} the algorithm uses the Clark Completion
resource rules for the resource predicates.  We obtain these rules by a 
preprocessing algorithm that needs to be done only once for a database.


\begin{description}
\item[Preprocessing Algorithm (Clark Completion)]
\item[Input:] ${\bf Res_{DB}}$. 
We may assume that each resource predicate $r$ is written in the form\\
$r(\bar{X}) \leftarrow p_1(\bar{X_1}),\ldots,p_n(\bar{X_n})$,\\
where $\bar{X}$ contains variables,
each $\bar{X_i}$ ($1 \leq i \leq n$)
consists of terms (constants or variables) and
$\bar{X} \subseteq \bigcup_{i=1}^{n} \bar{X_{i}}$.
\item[Output:] The Clark Completion resource rules {\bf CCrr} in clausal form.
\\
\\
\noindent
{\bf begin}
\item[Step 1.] Apply the Clark Completion to
each resource predicate definition to write it as\\
$r(\bar{X}) \leftrightarrow \exists \bar{Z} (p_1(\bar{X_1}), \ldots,
p_n(\bar{X_n}))$\\
where $\bar{Z}$ is the set of variables in
$(\bigcup_{i=1}^{n} \bar{X_{i}}) - \bar{X}$.
\item[Step 2.] Rewrite the equivalences obtained in Step 1 into rules
in clausal form, called the Clark Completion resource rules ({\bf CCrr}) as\\
$r(\bar{X}) \leftarrow p_1(\bar{X_1}), \ldots, p_n(\bar{X_n})$\\
$p_1(\bar{X'_1}) \leftarrow r(\bar{X})$\\
$\vdots$\\
$p_n(\bar{X'_n)}) \leftarrow r(\bar{X})$\\
where $\bar{X'_i}$ $(1 \leq i \leq n)$ is obtained from $\bar{X_i}$
by replacing every variable $X_j \in \bar{Z}$ by $f_{r,j}(\bar{X})$.
(In our examples we will usually use variables such as $X$, $Y$, and $Z$,
and function symbols $f$, $g$, and $h$, and omit subscripts.)
\end{description}
{\bf end}

We now describe the simplest form of the folding algorithm, the one with
the database restrictions of Section~\ref{restrict}.
%
%

The algorithm described below uses linear derivation (\cite{Chan73})
which includes a backtracking mechanism.  Backtracking occurs when we
find a linear derivation that has no resource predicates.
Before the algorithm commences we assume that there is a test to determine
which of the three cases applies to {\bf ICs}.  In Case 1, nothing has to be done.  In Case 2,
we assume that the linear derivation is modified to include a check to determine
if the clause, $L$, that has been generated, satisfies 
the bounding condition, and if it does, then backtracking occurs.  In Case 3, a 
depth bound $k$ is specified and if the depth is reached, backtracking occurs.
We also assume that if there are built-in predicates such as $=, \not=, \ge$,
then the input clauses are placed in expanded form.  If there are no built-in
predicates, it is unnecessary to do the expansion.

\begin{description}
\item[Folding Algorithm 1 (Finding a Single Folding)]
\item[Input:] $\cal C$: the set of clauses in the {\bf EDB}, {\bf IDB},
{\bf IC}, and {\bf CCrr}, and\\
the query, $q(\bar{X})$: $\leftarrow G(\bar{Y})$, where $\bar{X} \subseteq \bar{Y}$,
and $G$ is a conjunction of atoms.  We call $\bar{X}$ the query variables.
\item[Output:] A query $fq(\bar{X})$: $\leftarrow L(\bar{Z})$, where
$\bar{X} \subseteq \bar{Z}$. 
\\
\\
{\bf begin}\\
Starting with $\leftarrow G$ find a linear derivation
using $\cal C$ that results
in a clause $\leftarrow L$ that contains the query variables and no function symbols. When
$L$ contains at least one resource predicate,
and no {\bf EDB} predicates, it constitutes a complete folding;
otherwise $L$ may contain some {\bf EDB} predicates and
hence it constitutes a partial folding.
\\
{\bf end}
\end{description}

In our figures, we show illustrative derivations, but not the detailed steps
leading to that derivation that may have arisen by backtracking.
Starting with $\leftarrow G$, we essentially find a linear derivation 
using $\cal C$ that results
in a clause $\leftarrow L$ that contains the query variables.
If $L$ contains at least one resource predicate, 
and no {\bf EDB} predicates,
then it constitutes a complete folding;
otherwise $L$ may contain some {\bf EDB} predicates, then it
constitutes a partial folding.
Note that when the algorithm terminates with an answer,
backtracking may find additional answers.

At any point if an integrity constraint subsumes a clause in the
linear derivation, the process backs up because the query
cannot have any answers along that path.  We have omitted this 
step since subsumption is time consuming.  An algorithm for 
subsumption may be found in \cite{Chan73}.  If a clause in a derivation 
contains a constant, the expansion of a clause may allow us to derive
a solution.  This modification is omitted from the algorithm.


Next we show that the Folding Algorithm is {\em sound} or {\em correct}.  
By this we mean that every tuple obtained by solving the query $fq(\bar{X})$
is also obtained by solving the query $q(\bar{X})$.  That is, every answer 
to a folded query is an answer to the original query.

\begin{theorem}\label{sound}
The Folding Algorithm is sound (correct).
\end{theorem}

\begin{proof}
Using the notation $q(\bar{X}): \leftarrow G(\bar{Y})$ and 
$fq(\bar{X}): \leftarrow L(\bar{Z})$, by the soundness of 
resolution, we obtain\\
	 $(\leftarrow G(\bar{Y})) \cup {\cal C} \models (\leftarrow L(\bar{Z}))$,\\
so, by logical equivalence, we have,\\
	$L(\bar{Z}) \cup {\cal C} \models \leftarrow G(\bar{Y})$.\\
Suppose that $\bar{a}$ is a solution to $fq(\bar{X})$ in $DB$.  This means that
there is a $\bar{b}$, with $\bar{b}[\bar{X}] = \bar{a}$, such that $DB \models L(\bar{b})$.
Also, for every formula $C \in {\cal C}$, $DB \models C$.  Therefore, there is a
$\bar{d}$, where $\bar{d}[\bar{X}] = \bar{a}$, such that $DB \models G(\bar{d})$.
But this means that $\bar{a}$ is a solution to $q(\bar{X})$ in $DB$.
\end{proof}

\subsection{Additional Examples}

Next we apply our algorithm to various examples considered by researchers
and show that our algorithm can be used to obtain the same results.
We start by taking two examples from \cite{Q96}.
In that paper the {\bf EDB} consists of six relations that represent a
patient record database:

\begin{example}(Examples 2 and 3, Qian \cite{Q96})
\begin{tabbing}

{\bf Patients}\ \ \ \ \ \ \= (patient\_id,clinic,dob,insurance)\\
{\bf Physician} \>(physician\_id,clinic,pager\_no)\\
{\bf Drugs}             \>(drug\_name,generic?)\\
{\bf Notes}             \>(note\_id,patient\_id,physician\_id,note\_text)\\
{\bf Allergy}           \>(note\_id,drug\_name,allergy\_text)\\
{\bf Prescription}      \>(note\_id,drug\_name,prescription\_text)

\end{tabbing} 
The {\bf IDB} and {\bf IC} are empty; hence Case 1 of the {\bf IC} Cases (see page~\pageref{iccases}) applies.
The ${\bf Res_{DB}}$ consists of two relations, {\bf Drug\_Allergy} and 
{\bf Prescribed\_Drug}. 
For convenience we write {\bf Drug\_Allergy} as $r_1$ and
{\bf Prescribed-Drug} as $r_2$.
In {\bf Datalog} they are expressed as:
\[r_1(X_1, X_2, X_3) \leftarrow notes(U_1, X_1, U_2, U_3), 
allergy(U_1, X_2, X_3)\]
\[r_2(Y_1, Y_2, Y_3, Y_4) \leftarrow notes(V_1, Y_1, Y_2, V_2),
prescription(V_1, Y_3, V_3), drugs(Y_3, Y_4).\]

Preprocessing yields two Clark completion resource rules for $r_1$ and
three Clark completion resource rules for $r_2$.  In the following formulae 
the functions $f_i$, $g_j$ are abbreviations for the Skolem functions
$f_{i}(X_1,X_2,X_3)$ and $g_{j}(Y_1,Y_2,Y_3,Y_4)$, respectively.
\begin{equation}
CCrr1:  notes(f_1, X_1, f_2, f_3) \leftarrow r_1(X_1,X_2,X_3) \label{eq:11}
\end{equation}
\begin{equation}
CCrr2:  allergy(f_1, X_2, X_3) \leftarrow r_1(X_1,X_2,X_3) \label{eq:12}
\end{equation}
\begin{equation}
CCrr3:  notes(g_1,Y_1,Y_2,g_2) \leftarrow r_2(Y_1, Y_2, Y_3, Y_4) \label{eq:13}
\end{equation}
\begin{equation}
CCrr4:  prescription(g_1, Y_3, g_3) \leftarrow r_2(Y_1, Y_2, Y_3, Y_4) \label{eq:14}
\end{equation}
\begin{equation}
CCrr5:  drugs(Y_3,Y_4) \leftarrow r_2(Y_1, Y_2, Y_3, Y_4) \label{eq:15}
\end{equation}

Let's consider first the query of Examples 2 and 3 of \cite{Q96}:

\[q(X,Y): \leftarrow notes(W_1,X,W_2,W_3), allergy(W_1,Y,W_4),
notes(W_5,X,W_6,W_7), prescription(W_5,Y,W_8)\]

Again, as in our first example, we start with the body of the query
to find a derivation:
\[\leftarrow notes(W_1,X,W_2,W_3), allergy(W_1,Y,W_4), notes(W_5,X,W_6,W_7),
prescription(W_5,Y,W_8)\]
The derivation is shown in 
Figure 2.
Four of the Clark Completion resource rules are used.
The rewritten query at the bottom of the tree,\\
$\leftarrow r_1(X,Y,X_3), r_2(X,Y_2,Y,Y_4)$\\
consists of only resource predicates.

\end{example}

\begin{figure}
\label{qian2}
\setlength{\unitlength}{0.00083300in}%
\begingroup\makeatletter\ifx\SetFigFont\undefined
\def\x#1#2#3#4#5#6#7\relax{\def\x{#1#2#3#4#5#6}}%
\expandafter\x\fmtname xxxxxx\relax \def\y{splain}%
\ifx\x\y   
\gdef\SetFigFont#1#2#3{%
  \ifnum #1<17\tiny\else \ifnum #1<20\small\else
  \ifnum #1<24\normalsize\else \ifnum #1<29\large\else
  \ifnum #1<34\Large\else \ifnum #1<41\LARGE\else
     \huge\fi\fi\fi\fi\fi\fi
  \csname #3\endcsname}%
\else
\gdef\SetFigFont#1#2#3{\begingroup
  \count@#1\relax \ifnum 25<\count@\count@25\fi
  \def\x{\endgroup\@setsize\SetFigFont{#2pt}}%
  \expandafter\x
    \csname \romannumeral\the\count@ pt\expandafter\endcsname
    \csname @\romannumeral\the\count@ pt\endcsname
  \csname #3\endcsname}%
\fi
\fi\endgroup
\begin{picture}(5625,8817)(151,-8047)
\thicklines
\put(3601,539){\line( 0,-1){1200}}
\put(3601,-661){\line( 5, 2){2120.690}}
\put(3601,-961){\line( 0,-1){1200}}
\put(3601,-961){\line( 0,-1){1200}}
\put(3601,-2461){\line( 0,-1){900}}
\put(3601,-3661){\line( 0,-1){1200}}
\put(3605,-4870){\line( 5, 2){2120.690}}
\put(3601,-5236){\line( 0,-1){1200}}
\put(3617,-6397){\line( 5, 2){2120.690}}
\put(3601,-6811){\line( 0,-1){900}}
\put(3664,-2134){\line( 5, 2){2120.690}}
\put(151,614){\makebox(0,0)[lb]{\smash{\SetFigFont{12}{14.4}{rm}$Query: \leftarrow notes(W_1,X,W_2,W_3), allergy(W_1,Y,W_4), notes(W_5,X,W_6,W_7), prescription(W_5,Y,W_8)$}}}
\put(4876,-286){\makebox(0,0)[lb]{\smash{\SetFigFont{12}{14.4}{rm}$\{W_1/f_1, X_1/X, W_2/f_2, W_3/f_3\}$}}}
\put(1426,-886){\makebox(0,0)[lb]{\smash{\SetFigFont{12}{14.4}{rm}$ \leftarrow r_1(X,X_2,X_3), allergy(f_1,Y,W_4), notes(W_5,X,W_6,W_7), prescription(W_5,Y,W_8)$}}}
\put(3751,-2911){\makebox(0,0)[lb]{\smash{\SetFigFont{12}{14.4}{rm}factor}}}
\put(3751,-7261){\makebox(0,0)[lb]{\smash{\SetFigFont{12}{14.4}{rm}factor}}}
\put(4876,-6061){\makebox(0,0)[lb]{\smash{\SetFigFont{12}{14.4}{rm}$\{Y_3/Y,W_8/g_3\}$}}}
\put(4576,314){\makebox(0,0)[lb]{\smash{\SetFigFont{12}{14.4}{rm}$CCrr1: notes(f_1,X_1,f_2,f_3) \leftarrow r_1(X_1,X_2,X_3)$}}}
\put(4651,-3886){\makebox(0,0)[lb]{\smash{\SetFigFont{12}{14.4}{rm}$CCrr3: notes(g_1,Y_1,Y_2,g_2) \leftarrow r_2(Y_1,Y_2,Y_3,Y_4)$}}}
\put(4576,-5386){\makebox(0,0)[lb]{\smash{\SetFigFont{12}{14.4}{rm}$CCrr4: prescription(g_1,Y_3,g_3) \leftarrow r_2(Y_1,Y_2,Y_3,Y_4)$}}}
\put(4651,-1186){\makebox(0,0)[lb]{\smash{\SetFigFont{12}{14.4}{rm}$CCrr2: allergy(f_1,X_2,X_3) \leftarrow r_1(X_1,X_2,X_3)$}}}
\put(4876,-1786){\makebox(0,0)[lb]{\smash{\SetFigFont{12}{14.4}{rm}$\{X_2/Y, X_3 /W_4\}$}}}
\put(1426,-2386){\makebox(0,0)[lb]{\smash{\SetFigFont{12}{14.4}{rm}$ \leftarrow r_1(X,Y,W_4), r_1(X_1,Y,W_4), notes(W_5,X,W_6,W_7), prescription(W_5,Y,W_8)$}}}
\put(1426,-3586){\makebox(0,0)[lb]{\smash{\SetFigFont{12}{14.4}{rm}$ \leftarrow r_1(X,Y,W_4), notes(W_5,X,W_6,W_7), prescription(W_5,Y,W_8)$}}}
\put(4876,-4561){\makebox(0,0)[lb]{\smash{\SetFigFont{12}{14.4}{rm}$\{W_5/g_1, Y_1/X, Y_2/W_6, W_7/g_2\}$}}}
\put(1426,-5086){\makebox(0,0)[lb]{\smash{\SetFigFont{12}{14.4}{rm}$ \leftarrow r_1(X,Y,W_4), r_2(X,W_6,Y_3,Y_4), prescription(g_1,Y,W_8)$}}}
\put(1426,-6661){\makebox(0,0)[lb]{\smash{\SetFigFont{12}{14.4}{rm}$ \leftarrow r_1(X,Y,W_4), r_2(X,W_6,Y_3,Y_4), r_2(Y_1,W_6,Y,Y_4)$}}}
\put(1426,-8011){\makebox(0,0)[lb]{\smash{\SetFigFont{12}{14.4}{rm}$ \leftarrow r_1(X,Y,W_4), r_2(X,W_6,Y,Y_4)$}}}
\end{picture}
\caption{Example 2 from \cite{Q96}}
\end{figure}
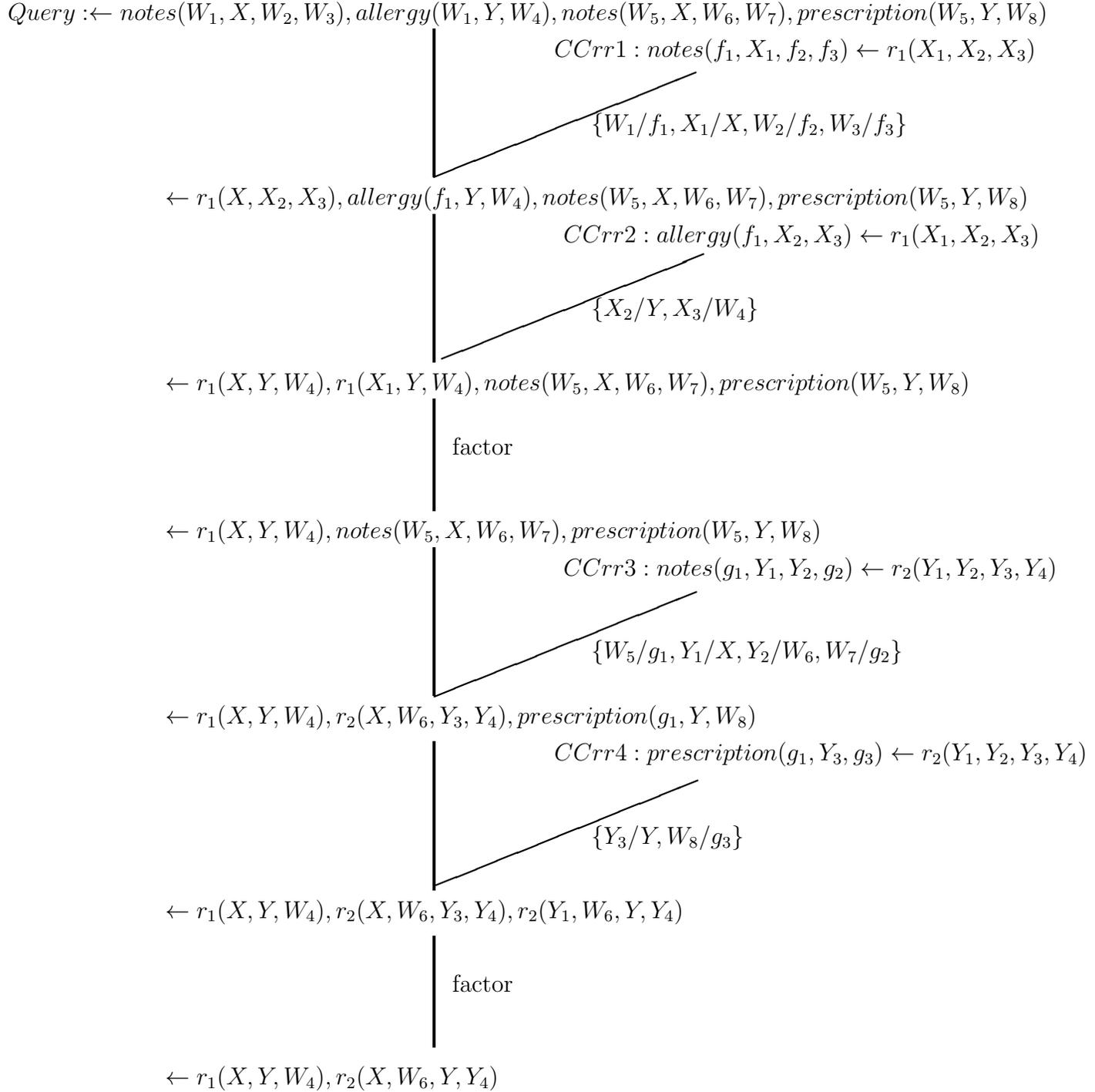

Now we consider Example 6 of \cite{Q96}.
\begin{example}(Example 6 of Qian \cite{Q96})
The {\bf EDB}, {\bf IDB} and {\bf IC} are the same as before.
${\bf Res_{DB}}$ consists of one relation $r$ defined as follows:
\[r(X_1, X_2, X_3) \leftarrow notes(U_1, X_1, X_2, U_2), 
prescription(U_1,X_3,U_3), drugs(X_3,U_4)\].

Preprocessing yields 
three Clark completion resource rules for $r$ as follows:
\begin{equation}
CCrr1:  notes(f_1, X_1, X_2, f_2) \leftarrow r(X_1,X_2,X_3) \label{eq:21}
\end{equation}
\begin{equation}
CCrr2:  prescription(f_1, X_3, f_3) \leftarrow r(X_1, X_2, X_3) \label{eq:22}
\end{equation}
\begin{equation}
CCrr3:  drugs(X_3,f_4) \leftarrow r(X_1, X_2, X_3) \label{eq:23}
\end{equation}

The query of this example is:

\[q(X,Y): \leftarrow patients(X,W_1,W_2,medicare), notes(W_3,X,W_4,W_5),
prescription(W_3,Y,W_6), drugs(Y,no)\]

For simplicity, we did not place the query in expanded form.
If we had, at the end we would have had to change the new 
variable back to the constant which it replaced.
The derivation is shown in 
Figure 3.
Two of the Clark completion resource rules are used.
The rewritten query at the bottom of the tree\\
\[\leftarrow patients(X,W_1,W_2,medicare), r(X,W_4,Y), drugs(Y,no)\]\\
consists of the resource predicate (replacing two extensional predicates)
and two extensional predicates.  This is an example of a partial folding.
\end{example}

\section{Handling Functional and Inclusion Dependencies}

\begin{figure} 
\label{qian6}
\setlength{\unitlength}{0.00083300in}%
\begingroup\makeatletter\ifx\SetFigFont\undefined
\def\x#1#2#3#4#5#6#7\relax{\def\x{#1#2#3#4#5#6}}%
\expandafter\x\fmtname xxxxxx\relax \def\y{splain}%
\ifx\x\y   
\gdef\SetFigFont#1#2#3{%
  \ifnum #1<17\tiny\else \ifnum #1<20\small\else
  \ifnum #1<24\normalsize\else \ifnum #1<29\large\else
  \ifnum #1<34\Large\else \ifnum #1<41\LARGE\else
     \huge\fi\fi\fi\fi\fi\fi
  \csname #3\endcsname}%
\else
\gdef\SetFigFont#1#2#3{\begingroup
  \count@#1\relax \ifnum 25<\count@\count@25\fi
  \def\x{\endgroup\@setsize\SetFigFont{#2pt}}%
  \expandafter\x
    \csname \romannumeral\the\count@ pt\expandafter\endcsname
    \csname @\romannumeral\the\count@ pt\endcsname
  \csname #3\endcsname}%
\fi
\fi\endgroup
\begin{picture}(5625,4470)(151,-3700)
\thicklines
\put(3601,539){\line( 0,-1){1200}}
\put(3601,-661){\line( 5, 2){2120.690}}
\put(3601,-961){\line( 0,-1){1200}}
\put(3664,-2134){\line( 5, 2){2120.690}}
\put(3601,-961){\line( 0,-1){1200}}
\put(3601,-2461){\line( 0,-1){900}}
\put(3751,-2911){\makebox(0,0)[lb]{\smash{\SetFigFont{12}{14.4}{rm}factor}}}
\put(4576,314){\makebox(0,0)[lb]{\smash{\SetFigFont{12}{14.4}{rm}$CCrr1: notes(f_1,X_1,f_2,f_3) \leftarrow r_1(X_1,X_2,X_3)$}}}
\put(151,614){\makebox(0,0)[lb]{\smash{\SetFigFont{12}{14.4}{rm}$Query: \leftarrow  patients(X,W_1,W_2,medicare), notes(W_3,X,W_4,W_5), prescription(W_3,Y,W_6), drugs(Y,no)$}}}
\put(4876,-286){\makebox(0,0)[lb]{\smash{\SetFigFont{12}{14.4}{rm}$\{W_3/f_1, X_1/X, X_2/W_4, W_5/f_2\}$}}}
\put(4651,-1186){\makebox(0,0)[lb]{\smash{\SetFigFont{12}{14.4}{rm}$CCrr2: prescription(f_1,X_3,f_3) \leftarrow r_1(X_1,X_2,X_3)$}}}
\put(4876,-1786){\makebox(0,0)[lb]{\smash{\SetFigFont{12}{14.4}{rm}$\{X_3/Y, W_6/f_3\}$}}}
\put(1426,-811){\makebox(0,0)[lb]{\smash{\SetFigFont{12}{14.4}{rm}$ \leftarrow  patients(X,W_1,W_2,medicare),  r(X,W_4,X_3), prescription(f_1,Y,W_6), drugs(Y,no)$}}}
\put(1426,-2311){\makebox(0,0)[lb]{\smash{\SetFigFont{12}{14.4}{rm}$ \leftarrow  patients(X,W_1,W_2,medicare),  r(X,W_4,Y), r(X_1,X_2,Y), drugs(Y,no)$}}}
\put(1426,-2311){\makebox(0,0)[lb]{\smash{\SetFigFont{12}{14.4}{rm}$ \leftarrow  patients(X,W_1,W_2,medicare),  r(X,W_4,Y), r(X_1,X_2,Y), drugs(Y,no)$}}}
\put(1426,-2311){\makebox(0,0)[lb]{\smash{\SetFigFont{12}{14.4}{rm}$ \leftarrow  patients(X,W_1,W_2,medicare),  r(X,W_4,Y), r(X_1,X_2,Y), drugs(Y,no)$}}}
\put(1426,-2311){\makebox(0,0)[lb]{\smash{\SetFigFont{12}{14.4}{rm}$ \leftarrow  patients(X,W_1,W_2,medicare),  r(X,W_4,Y), r(X_1,X_2,Y), drugs(Y,no)$}}}
\put(1426,-2311){\makebox(0,0)[lb]{\smash{\SetFigFont{12}{14.4}{rm}$ \leftarrow  patients(X,W_1,W_2,medicare),  r(X,W_4,Y), r(X_1,X_2,Y), drugs(Y,no)$}}}
\put(1426,-2311){\makebox(0,0)[lb]{\smash{\SetFigFont{12}{14.4}{rm}$ \leftarrow  patients(X,W_1,W_2,medicare),  r(X,W_4,Y), r(X_1,X_2,Y), drugs(Y,no)$}}}
\put(1426,-2311){\makebox(0,0)[lb]{\smash{\SetFigFont{12}{14.4}{rm}$ \leftarrow  patients(X,W_1,W_2,medicare),  r(X,W_4,Y), r(X_1,X_2,Y), drugs(Y,no)$}}}
\put(1426,-2311){\makebox(0,0)[lb]{\smash{\SetFigFont{12}{14.4}{rm}$ \leftarrow  patients(X,W_1,W_2,medicare),  r(X,W_4,Y), r(X_1,X_2,Y), drugs(Y,no)$}}}
\put(1426,-2311){\makebox(0,0)[lb]{\smash{\SetFigFont{12}{14.4}{rm}$ \leftarrow  patients(X,W_1,W_2,medicare),  r(X,W_4,Y), r(X_1,X_2,Y), drugs(Y,no)$}}}
\put(1426,-2311){\makebox(0,0)[lb]{\smash{\SetFigFont{12}{14.4}{rm}$ \leftarrow  patients(X,W_1,W_2,medicare),  r(X,W_4,Y), r(X_1,X_2,Y), drugs(Y,no)$}}}
\put(1426,-2311){\makebox(0,0)[lb]{\smash{\SetFigFont{12}{14.4}{rm}$ \leftarrow  patients(X,W_1,W_2,medicare),  r(X,W_4,Y), r(X_1,X_2,Y), drugs(Y,no)$}}}
\put(1426,-3661){\makebox(0,0)[lb]{\smash{\SetFigFont{12}{14.4}{rm}$ \leftarrow  patients(X,W_1,W_2,medicare),  r(X,W_4,Y), drugs(Y,no)$}}}
\end{picture}
\caption{Example 6 from \cite{Q96}}
\end{figure}
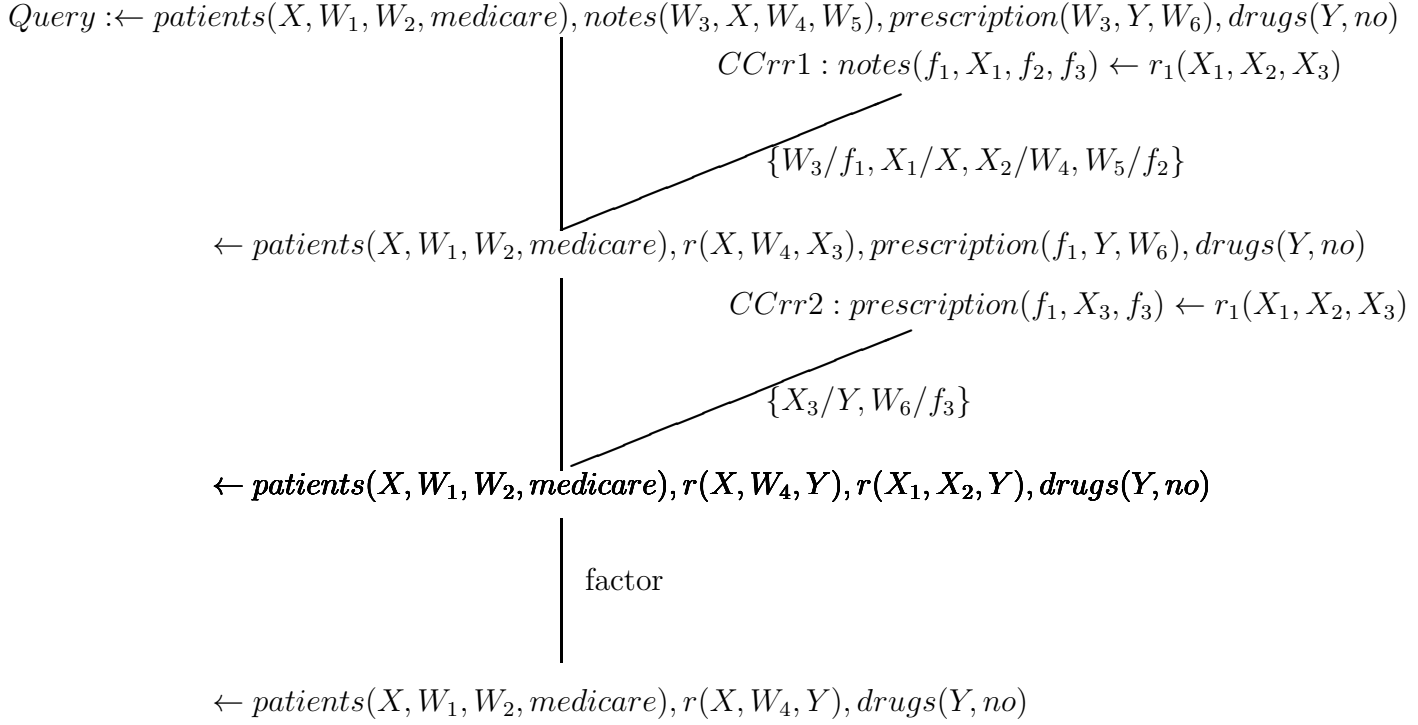 

This section illustrates the use of our algorithm in the special cases
where the integrity constraints are functional and inclusion dependencies.
As explained earlier, the presence of functional dependencies means that
Case 3 for {\bf ICs} applies (see page~\pageref{iccases}).
\cite{DGQ96} gives algorithms in the presence of functional dependencies.
Their basic idea is to use a functional dependency to decompose a relation
into several relations, using a lossless join decomposition, and then
apply the standard folding algorithm.
We show that such a decomposition is not necessary.
Instead of decomposing a relation that contains a functional dependency,
we generate additional Clark completion rules and then apply our standard
algorithm.

\subsection{Example with Key Constraint}

We consider how our algorithm applies to Example 3 of \cite{DGQ96}.
The {\bf EDB} contains three relations, again involving a patient
record database as follows:  

\begin{tabbing}

{\bf Patients}\ \ \ \ \ \ \= (name,dob,insurance)\\
{\bf Procedure}     \>(patient\_name,physician\_name,procedure\_name,time)\\
{\bf Insurer}       \>(company,address,phone)\\

\end{tabbing} 
Again, {\bf IDB} = $\emptyset$.
The ${\bf Res_{DB}}$ consists of two relations 
{\bf Clinical\_History} and {\bf Billing} which we write as $r_1$ 
and $r_2$ with the following definitions:
\[r_1(X_1, X_2, X_3, X_4) \leftarrow patients(X_1,X_2,U_1,U_2),
procedure(X_1,U_3,X_3,X_4)\]
\[r_2(Y_1,Y_2,Y_3) \leftarrow patients(Y_1,V_1,Y_2,V_2), insurer(V_2,Y_3,V_3)\]

The {\bf IC} contains the key constraint  $patients: patient\_id
\rightarrow clinic, dob, insurance$, written in Datalog as three clauses:\\
$X_2 = Y_2 \leftarrow patients(X_1,X_2,X_3,X_4), patients(X_1,Y_2,Y_3,Y_4)$\\
$X_3 = Y_3 \leftarrow patients(X_1,X_2,X_3,X_4), patients(X_1,Y_2,Y_3,Y_4)$\\
$X_4 = Y_4 \leftarrow patients(X_1,X_2,X_3,X_4), patients(X_1,Y_2,Y_3,Y_4)$\\

Preprocessing yields 
four Clark completion resource rules for $r_1$ and $r_2$ as follows:
\begin{equation}
CCrr1:  patients(X_1, X_2, f_1, f_2) \leftarrow r_1(X_1,X_2,X_3,X_4) \label{eq:31}
\end{equation}
\begin{equation}
CCrr2:  procedure(X_1, f_2, X_3, X_4) \leftarrow r_1(X_1, X_2, X_3, X_4) \label{eq:32}
\end{equation}
\begin{equation}
CCrr3:  patients(Y_1,g_1,Y_2,g_2) \leftarrow r_2(Y_1, Y_2, Y_3) \label{eq:33}
\end{equation}
\begin{equation} 
CCrr4:  insurer(g_2,Y_3,g_3) \leftarrow r_2(Y_1, Y_2, Y_3) \label{eq:34}
\end{equation}

As we will show below, in the case of a key constraint it is sometimes possible
to combine Clark completion resource rules. In this particular case the
rules Equation~\ref{eq:31} and Equation~\ref{eq:33} can be combined to yield
\begin{equation}
Combined CCrr: patients(X_1,X_2,Y_2,f_2) \leftarrow 
r_1(X_1,X_2,X_3,X_4),r_2(X_1,Y_2,Y_3)
\label{eq:35}
\end{equation} 

The query of this example is:

\[q(X,Y,Z): \leftarrow patients(X,Y,Z,W)\]

\begin{figure}
\label{key-constraint}
\setlength{\unitlength}{0.00083300in}%
\begingroup\makeatletter\ifx\SetFigFont\undefined
\def\x#1#2#3#4#5#6#7\relax{\def\x{#1#2#3#4#5#6}}%
\expandafter\x\fmtname xxxxxx\relax \def\y{splain}%
\ifx\x\y   
\gdef\SetFigFont#1#2#3{%
  \ifnum #1<17\tiny\else \ifnum #1<20\small\else
  \ifnum #1<24\normalsize\else \ifnum #1<29\large\else
  \ifnum #1<34\Large\else \ifnum #1<41\LARGE\else
     \huge\fi\fi\fi\fi\fi\fi
  \csname #3\endcsname}%
\else
\gdef\SetFigFont#1#2#3{\begingroup
  \count@#1\relax \ifnum 25<\count@\count@25\fi
  \def\x{\endgroup\@setsize\SetFigFont{#2pt}}%
  \expandafter\x
    \csname \romannumeral\the\count@ pt\expandafter\endcsname
    \csname @\romannumeral\the\count@ pt\endcsname
  \csname #3\endcsname}%
\fi
\fi\endgroup
\begin{picture}(3225,2595)(1426,-2425)
\thicklines
\put(3001,-136){\line( 0,-1){1950}}
\put(3064,-1995){\line( 4, 3){1584}}
\put(1576, 14){\makebox(0,0)[lb]{\smash{\SetFigFont{12}{14.4}{rm}$Query:  \leftarrow patients(X,Y,Z,W)$}}}
\put(1426,-2386){\makebox(0,0)[lb]{\smash{\SetFigFont{12}{14.4}{rm}$\leftarrow r_{1}(X,Y,X_{3},X_{4}), r_{2}(X,Z,Y_{3})$}}}
\put(3301,-661){\makebox(0,0)[lb]{\smash{\SetFigFont{12}{14.4}{rm}$Combined CCr: patients(X_1,X_2,Y_2,f_2) \leftarrow r_1(X_1,X_2,X_3,X_4), r_2(X_1,Y_2,Y_3)$}}}
\put(3976,-1486){\makebox(0,0)[lb]{\smash{\SetFigFont{12}{14.4}{rm}$\{X_1/X,X_2/Y,Y_2/Z,W/f_2\}$}}}
\end{picture}
\caption{Key Constraint Example}
\end{figure}
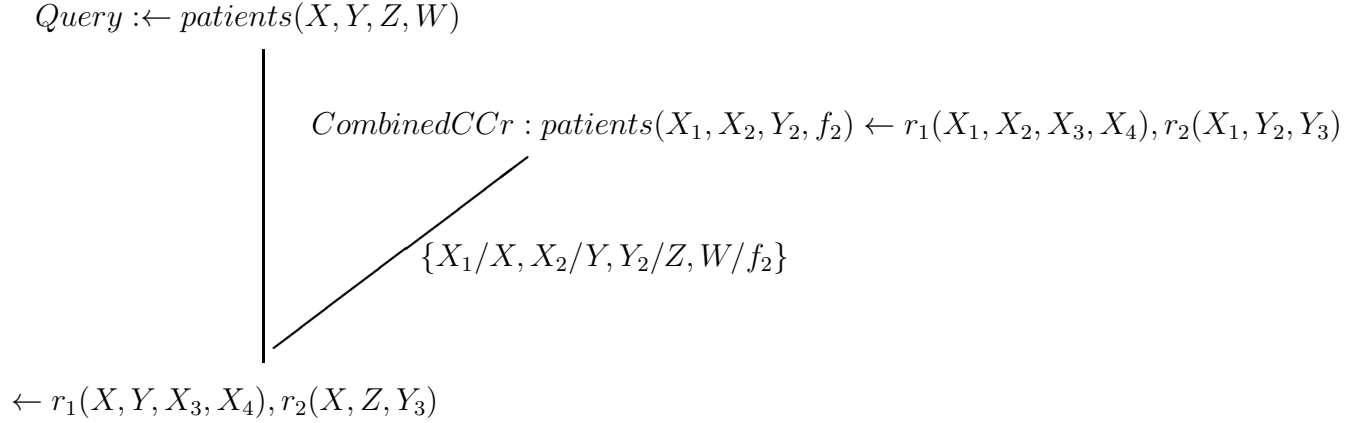

As shown in 
Figure 4, starting with the body of this query and using the combined
Clark Completion resource rule, the derivation takes one step to obtain\\
\[\leftarrow r_1(X,Y,X_3,X_4),r_2(X,Z,Y_3)\]\\
We note, however, that we could not answer the query 

\[q(X,Y,Z,W): \leftarrow patients(X,Y,Z,W)\]\\
this way because $W$ does not appear in the folded query.

\subsection{Combining Clark Completion Resource Rules}

In the presence of functional dependencies it is possible under certain 
conditions to combine Clark completion resource rules for the same predicate 
in such a way that function symbols are replaced by variables.  Using the 
combined rules may simplify the derivation of the folded query.  In this
subsection we deal with the special useful case of key constraints.

We start by introducing notation involved in combining two Clark completion
resource rules.  The same basic method as described below will handle more
than two rules.   We assume that there exist two such rules of the form
\begin{equation}
p(\bar{Xf}) \leftarrow r_1(\bar{X}) \label{eq:41}
\end{equation}
\begin{equation}
p(\bar{Yg}) \leftarrow r_2(\bar{Y}) \label{eq:42}
\end{equation}

where $p$ is an n-ary predicate, $\bar{Xf}$ is an n-tuple all of whose variables are
in $\bar{X}$ and may contain functions symbols $f_i$, and $\bar{Yg}$ is an n-tuple 
all of whose variables are in $\bar{Y}$ and may contain functions symbols $g_j$.
For any tuple $\bar{U}$ we write $\bar{U}[i]$ for the i-th component of $\bar{U}$.
Define the n-ary tuple $\bar{Z}$

\[ \bar{Z}[i] = \left\{ \begin{array}{ll}
                        \bar{Yg}[i] & \mbox{if $\bar{Xf}[i]$ is a function symbol and 
                          $\bar{Yg}[i]$ is a variable} \\
                        \bar{Xf}[i] & \mbox{otherwise}
                      \end{array}
                \right. \]

Also define the tuple $\bar{Y}/(\bar{X}k)$ to have the same number of components as 
$\bar{Y}$ and defined as 

\[ \bar{Y}/(\bar{X}k)[i] = \left\{ \begin{array}{ll}
                        \bar{X}[i] & \mbox{if $1 \leq i \leq k $} \\
                        \bar{Y}[i] & \mbox{if $k < i$} 
                      \end{array}
                \right. \]

\newtheorem{guess}{Proposition}
\begin{guess}
For the two rules given in \ref{eq:41} and \ref{eq:42}, 
if $\bar{Xf}[i]$ and  $\bar{Yg}[i]$ are variables for $1 \leq i \leq k$ and the
first $k$ columns of $p$ form a key for $p$, then
the combined Clark completion resource rule written as
\begin{equation}
p(\bar{Z}) \leftarrow r_1(\bar{X}), r_2(\bar{Y}/(\bar{X}k))
\end{equation}
is also a valid rule.
\end{guess}
\begin{proof}
Proof: By~\ref{eq:41}
\begin{equation}
p(\bar{Xf}) \leftarrow r_1(\bar{X}), r_2(\bar{Y}/(\bar{X}k))
\end{equation}
By \ref{eq:42}
\begin{equation}
p(\bar{Yg}/(\bar{X}k)) \leftarrow r_1(\bar{X}), r_2(\bar{Y}/(\bar{X}k)).
\end{equation}
By the hypothesis that $\bar{Xf}[i]$ and $\bar{Yg}[i]$ are variables for
$1 \leq i \leq k$, we obtain $\bar{Xf}[i] = \bar{Yg}/(\bar{X}k)$ for 
$1 \leq i \leq k$.  As the first $k$ columns of $p$ form a key, the
corresponding elements of $\bar{Xf}$ and $\bar{Yg}/(\bar{X}k)$ must be
equal.  Since $\bar{Z}$ contains the first $k$ columns of $\bar{Xf}$
and the rest of the columns are from $\bar{Xf}$ or $\bar{Yg}$, 
\begin{equation}
p(\bar{Z}) \leftarrow r_1(\bar{X}), r_2(\bar{Y}/(\bar{X}k))
\end{equation}
follows.
\end{proof}

Going back to the example of Section 4.1,
$CCrr1$ and $CCrr3$ are two Clark completion resource rules 
for the $patients$
predicate.
The first attribute is the key and both have a variable for the first
attribute: $X_1$ and $Y_1$.
Now set  $Y_1 = X_1$.
This forces  $g_1 = X_2$, $f_1 = Y_2$, and $g_2 = f_2$, and we obtain the
$Combined~CCrr$.
\subsection{Using the Lossless Join Decomposition Property for Key Constraints}

In \cite{DGQ96} the query in 4.1 is solved using a property of key 
constraints.
Namely, the key constraint  $patients: patient\_id \rightarrow 
clinic, dob, insurance$ implies that the decomposition of the relation
$patients(patient\_id,clinic,dob,insurance)$ to the three relations
$pat1(patient\_id,clinic)$, $pat2(patient\_id,dob)$, 
$pat3(patient\_id,insurance)$ is a lossless join decomposition.
Therefore we can deal with the three relations $pat1$, $pat2$, $pat3$
instead of $patients$.

Now, the {\bf Res_{DB}} relations are defined as follows:\\
\begin{equation}
r_1(X_1,X_2,X_3,X_4) \leftarrow pat1(X_1,X_2), pat2(X_1,U_1),
pat3(X_1,U_2), procedure(X_1,U_3,X_3,X_4)
\end{equation}
\begin{equation}
r_2(Y_1,Y_2,Y_3) \leftarrow pat1(Y_1,V_1), pat2(Y_1,Y_2),
pat3(Y_1,V_2), insurer(V_2,Y_3,V_3)
\end{equation}
and there are eight Clark Completion resource rules:
\begin{equation}
CCrr1:  pat1(X_1,X_2) \leftarrow r_1(X_1,X_2,X_3,X_4)
\end{equation}
\begin{equation}
CCrr2:  pat2(X_1,f_1) \leftarrow r_1(X_1,X_2,X_3,X_4)
\end{equation}
\begin{equation}
CCrr3:  pat3(X_1,f_2) \leftarrow r_1(X_1,X_2,X_3,X_4)
\end{equation}
\begin{equation}
CCrr4:  procedure(X_1,f_3,X_3,X_4) \leftarrow r_1(X_1,X_2,X_3,X_4)
\end{equation}
\begin{equation}
CCrr5:  pat1(Y_1,g_1) \leftarrow r_2(Y_1,Y_2,Y_3)
\end{equation}
\begin{equation}
CCrr6:  pat2(Y_1,Y_2) \leftarrow r_2(Y_1,Y_2,Y_3)
\end{equation}
\begin{equation}
CCrr7:  pat3(Y_1,g_2) \leftarrow r_2(Y_1,Y_2,Y_3)
\end{equation}
\begin{equation}
CCrr8:  insurer(g_2,Y_3,g_3) \leftarrow r_2(Y_1,Y_2,Y_3)
\end{equation}

The query
\begin{equation}
q(X,Y,Z): \leftarrow patients(X,Y,Z,W)
\end{equation}
is rewritten as
\begin{equation}
q(X,Y,Z): \leftarrow pat1(X,Y), pat2(X,Z), pat3(X,W)
\end{equation}
but $pat3(X,W)$ is superfluous, because the query does not 
contain $W$, hence we obtain
\begin{equation}
q(X,Y,Z): \leftarrow pat1(X,Y), pat2(X,Z)
\end{equation}
The derivation is shown in 
Figure 5.
Two of the eight Clark Completion resource rules are used.
The final query is the same as the one we obtained using the
Combined Clark Completion rule.
Again, if the query were
\begin{equation}
q(X,Y,Z,W): \leftarrow patients(X,Y,Z,W)
\end{equation}
then we could not obtain a complete folding because after applying
$CCrr1$ and $CCrr6$ we would be left with
\begin{equation}
\leftarrow r_1(X,Y,X_3,X_4), r_2(X,Z,Y_3),pat3(X,Y)
\end{equation}
and now applying applying either $CCrr3$ or $CCrr7$ would lead to a
function symbol for one of the variables in $r_1$ or $r_2$.

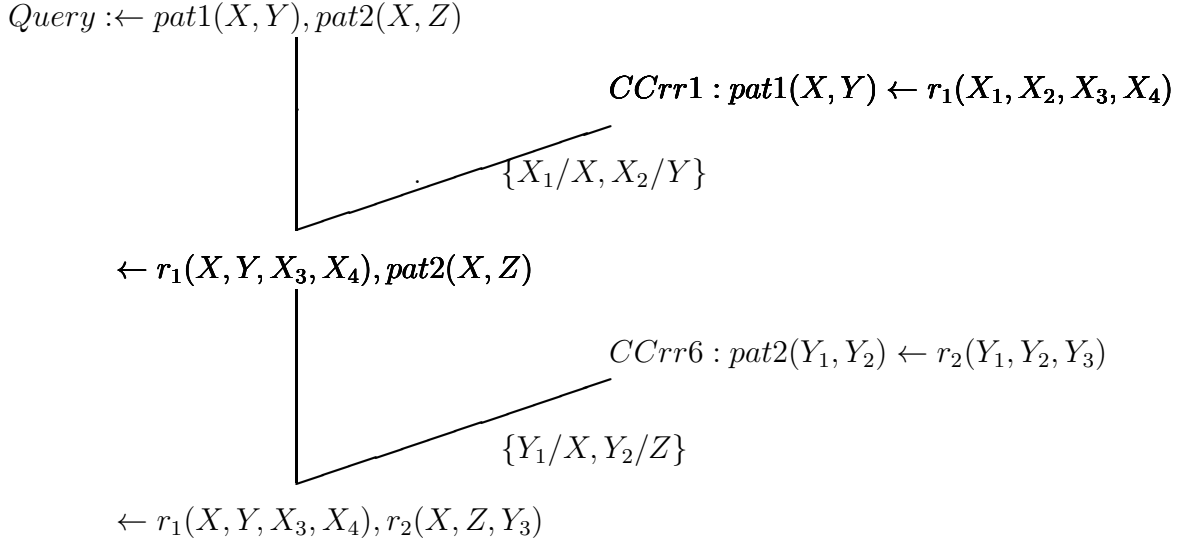
\begin{figure}
\label{decomp}
\setlength{\unitlength}{0.00083300in}%
\begingroup\makeatletter\ifx\SetFigFont\undefined
\def\x#1#2#3#4#5#6#7\relax{\def\x{#1#2#3#4#5#6}}%
\expandafter\x\fmtname xxxxxx\relax \def\y{splain}%
\ifx\x\y   
\gdef\SetFigFont#1#2#3{%
  \ifnum #1<17\tiny\else \ifnum #1<20\small\else
  \ifnum #1<24\normalsize\else \ifnum #1<29\large\else
  \ifnum #1<34\Large\else \ifnum #1<41\LARGE\else
     \huge\fi\fi\fi\fi\fi\fi
  \csname #3\endcsname}%
\else
\gdef\SetFigFont#1#2#3{\begingroup
  \count@#1\relax \ifnum 25<\count@\count@25\fi
  \def\x{\endgroup\@setsize\SetFigFont{#2pt}}%
  \expandafter\x
    \csname \romannumeral\the\count@ pt\expandafter\endcsname
    \csname @\romannumeral\the\count@ pt\endcsname
  \csname #3\endcsname}%
\fi
\fi\endgroup
\begin{picture}(3762,3345)(2401,-2950)
\thicklines
\put(4201,164){\line( 0,-1){1200}}
\put(4201,-1036){\line( 3, 1){1957.500}}
\put(4201,-286){\makebox(6.6667,10.0000){\SetFigFont{10}{12}{rm}.}}
\put(4951,-736){\makebox(6.6667,10.0000){\SetFigFont{10}{12}{rm}.}}
\put(4201,-286){\makebox(6.6667,10.0000){\SetFigFont{10}{12}{rm}.}}
\put(4201,-1411){\line( 0,-1){1200}}
\put(4201,-2621){\line( 3, 1){1957.500}}
\put(2401,239){\makebox(0,0)[lb]{\smash{\SetFigFont{12}{14.4}{rm}$Query: \leftarrow pat1(X,Y), pat2(X,Z)$}}}
\put(6151,-211){\makebox(0,0)[lb]{\smash{\SetFigFont{12}{14.4}{rm}$CCrr1: pat1(X,Y) \leftarrow r_1(X_1,X_2,X_3,X_4)$}}}
\put(5476,-736){\makebox(0,0)[lb]{\smash{\SetFigFont{12}{14.4}{rm}$\{X_1/X, X_2/Y\}$}}}
\put(3076,-1336){\makebox(0,0)[lb]{\smash{\SetFigFont{12}{14.4}{rm}$\leftarrow r_1(X,Y,X_3,X_4), pat2(X,Z)$}}}
\put(6151,-211){\makebox(0,0)[lb]{\smash{\SetFigFont{12}{14.4}{rm}$CCrr1: pat1(X,Y) \leftarrow r_1(X_1,X_2,X_3,X_4)$}}}
\put(6151,-211){\makebox(0,0)[lb]{\smash{\SetFigFont{12}{14.4}{rm}$CCrr1: pat1(X,Y) \leftarrow r_1(X_1,X_2,X_3,X_4)$}}}
\put(6151,-211){\makebox(0,0)[lb]{\smash{\SetFigFont{12}{14.4}{rm}$CCrr1: pat1(X,Y) \leftarrow r_1(X_1,X_2,X_3,X_4)$}}}
\put(6151,-211){\makebox(0,0)[lb]{\smash{\SetFigFont{12}{14.4}{rm}$CCrr1: pat1(X,Y) \leftarrow r_1(X_1,X_2,X_3,X_4)$}}}
\put(6151,-211){\makebox(0,0)[lb]{\smash{\SetFigFont{12}{14.4}{rm}$CCrr1: pat1(X,Y) \leftarrow r_1(X_1,X_2,X_3,X_4)$}}}
\put(6151,-211){\makebox(0,0)[lb]{\smash{\SetFigFont{12}{14.4}{rm}$CCrr1: pat1(X,Y) \leftarrow r_1(X_1,X_2,X_3,X_4)$}}}
\put(6151,-211){\makebox(0,0)[lb]{\smash{\SetFigFont{12}{14.4}{rm}$CCrr1: pat1(X,Y) \leftarrow r_1(X_1,X_2,X_3,X_4)$}}}
\put(6151,-211){\makebox(0,0)[lb]{\smash{\SetFigFont{12}{14.4}{rm}$CCrr1: pat1(X,Y) \leftarrow r_1(X_1,X_2,X_3,X_4)$}}}
\put(6151,-211){\makebox(0,0)[lb]{\smash{\SetFigFont{12}{14.4}{rm}$CCrr1: pat1(X,Y) \leftarrow r_1(X_1,X_2,X_3,X_4)$}}}
\put(6151,-211){\makebox(0,0)[lb]{\smash{\SetFigFont{12}{14.4}{rm}$CCrr1: pat1(X,Y) \leftarrow r_1(X_1,X_2,X_3,X_4)$}}}
\put(6151,-1861){\makebox(0,0)[lb]{\smash{\SetFigFont{12}{14.4}{rm}$CCrr6: pat2(Y_1,Y_2) \leftarrow r_2(Y_1,Y_2,Y_3)$}}}
\put(5476,-2461){\makebox(0,0)[lb]{\smash{\SetFigFont{12}{14.4}{rm}$\{Y_1/X, Y_2/Z\}$}}}
\put(3076,-1336){\makebox(0,0)[lb]{\smash{\SetFigFont{12}{14.4}{rm}$\leftarrow r_1(X,Y,X_3,X_4), pat2(X,Z)$}}}
\put(3076,-1336){\makebox(0,0)[lb]{\smash{\SetFigFont{12}{14.4}{rm}$\leftarrow r_1(X,Y,X_3,X_4), pat2(X,Z)$}}}
\put(3076,-1336){\makebox(0,0)[lb]{\smash{\SetFigFont{12}{14.4}{rm}$\leftarrow r_1(X,Y,X_3,X_4), pat2(X,Z)$}}}
\put(3076,-1336){\makebox(0,0)[lb]{\smash{\SetFigFont{12}{14.4}{rm}$\leftarrow r_1(X,Y,X_3,X_4), pat2(X,Z)$}}}
\put(3076,-1336){\makebox(0,0)[lb]{\smash{\SetFigFont{12}{14.4}{rm}$\leftarrow r_1(X,Y,X_3,X_4), pat2(X,Z)$}}}
\put(3076,-1336){\makebox(0,0)[lb]{\smash{\SetFigFont{12}{14.4}{rm}$\leftarrow r_1(X,Y,X_3,X_4), pat2(X,Z)$}}}
\put(3076,-2911){\makebox(0,0)[lb]{\smash{\SetFigFont{12}{14.4}{rm}$\leftarrow r_1(X,Y,X_3,X_4), r_2(X,Z,Y_3)$}}}
\end{picture}
\caption{Key Constraint Example Using Relation Decomposition}
\end{figure}

\subsection{Decomposition Cannot Always Handle Functional Dependencies}

Our next example also contains a functional dependency, but in this case
the functional dependency cannot be handled by a decomposition.
However, our algorithm can be used to obtain a folding.
The {\bf EDB} consists of two relations\\
${\bf p_1}(X,Y)$  and  ${\bf p_2}(X,Y)$\\
the {\bf IDB} is empty and the {\bf IC} contains the key constraint
$p_2: X \rightarrow Y$, written as:\\
$Y = Y' \leftarrow p_2(X',Y), p_2(X',Y')$.\\
Note that neither {\bf EDB} relation can be decomposed.\\
The {\bf Res_{DB}} consists of two relations:
\begin{equation}
r_1(X,Z) \leftarrow p_1(X,W), p_2(Z,W)
\end{equation}
\begin{equation}
r_2(X,Y) \leftarrow p_2(X,Y)
\end{equation}
Preprocessing yields two Clark completion resource rules for $r_1$
and one Clark completion resource rule for $r_2$ as follows:
\begin{equation}
p_1(X,f) \leftarrow r_1(X,Z)
\end{equation}
\begin{equation}
p_2(Z,f) \leftarrow r_1(X,Z)
\end{equation}
\begin{equation}
p_2(X,Y) \leftarrow r_2(X,Y)
\end{equation}
Consider the query
\begin{equation}
q(X): \leftarrow p_1(X,c)
\end{equation}
The solution is given in 
Figure 6.
We start by expanding the query, that is, rewriting the query to
\begin{equation}
q(X): \leftarrow p_1(X,W), W = c
\end{equation}
in order to take the constant out of the predicate allowing for a
substitution later.
All three Clark completion resource rules are used in the derivation
to obtain
\begin{equation}
\leftarrow r_1(X,Z), r_2(Z,c)
\end{equation}
which is a complete folding.

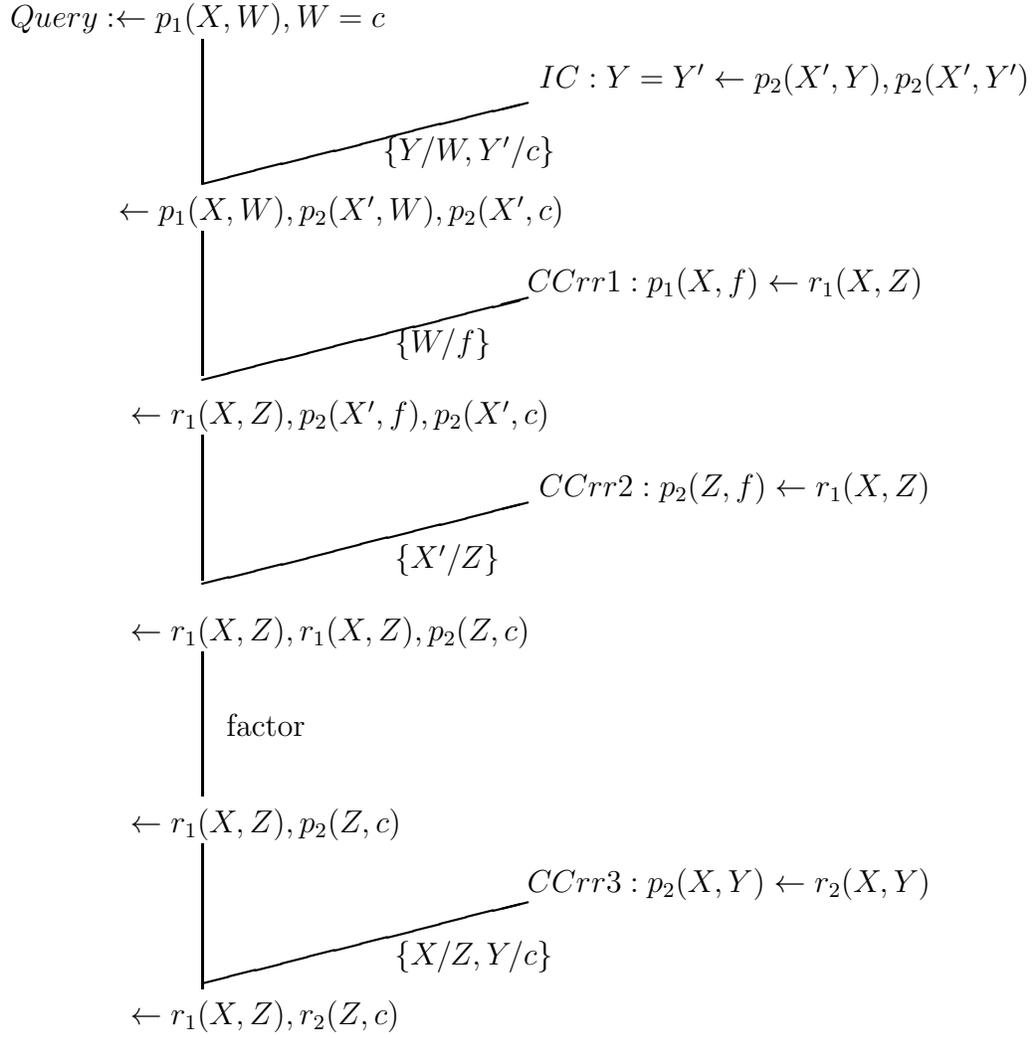
\begin{figure}
\label{fd-example}
\setlength{\unitlength}{0.00083300in}%
\begingroup\makeatletter\ifx\SetFigFont\undefined
\def\x#1#2#3#4#5#6#7\relax{\def\x{#1#2#3#4#5#6}}%
\expandafter\x\fmtname xxxxxx\relax \def\y{splain}%
\ifx\x\y   
\gdef\SetFigFont#1#2#3{%
  \ifnum #1<17\tiny\else \ifnum #1<20\small\else
  \ifnum #1<24\normalsize\else \ifnum #1<29\large\else
  \ifnum #1<34\Large\else \ifnum #1<41\LARGE\else
     \huge\fi\fi\fi\fi\fi\fi
  \csname #3\endcsname}%
\else
\gdef\SetFigFont#1#2#3{\begingroup
  \count@#1\relax \ifnum 25<\count@\count@25\fi
  \def\x{\endgroup\@setsize\SetFigFont{#2pt}}%
  \expandafter\x
    \csname \romannumeral\the\count@ pt\expandafter\endcsname
    \csname @\romannumeral\the\count@ pt\endcsname
  \csname #3\endcsname}%
\fi
\fi\endgroup
\begin{picture}(3300,6420)(1201,-5875)
\thicklines
\put(2401,314){\line( 0,-1){900}}
\put(2401,-586){\line( 4, 1){2029.412}}
\put(2401,-886){\line( 0,-1){900}}
\put(2401,-1809){\line( 4, 1){2029.412}}
\put(2401,-2161){\line( 0,-1){900}}
\put(2401,-3086){\line( 4, 1){2029.412}}
\put(2401,-3511){\line( 0,-1){900}}
\put(2401,-4711){\line( 0,-1){900}}
\put(2401,-5579){\line( 4, 1){2029.412}}
\put(1201,389){\makebox(0,0)[lb]{\smash{\SetFigFont{12}{14.4}{rm}$Query: \leftarrow p_1(X,W), W = c$}}}
\put(4501, 14){\makebox(0,0)[lb]{\smash{\SetFigFont{12}{14.4}{rm}$IC: Y = Y' \leftarrow p_2(X',Y), p_2(X',Y')$}}}
\put(1876,-811){\makebox(0,0)[lb]{\smash{\SetFigFont{12}{14.4}{rm}$\leftarrow p_1(X,W), p_2(X',W), p_2(X',c)$}}}
\put(4426,-1261){\makebox(0,0)[lb]{\smash{\SetFigFont{12}{14.4}{rm}$CCrr1: p_1(X,f) \leftarrow r_1(X,Z)$}}}
\put(3601,-1636){\makebox(0,0)[lb]{\smash{\SetFigFont{12}{14.4}{rm}$\{W/f\}$}}}
\put(1951,-2086){\makebox(0,0)[lb]{\smash{\SetFigFont{12}{14.4}{rm}$\leftarrow r_1(X,Z), p_2(X',f), p_2(X',c)$}}}
\put(4501,-2536){\makebox(0,0)[lb]{\smash{\SetFigFont{12}{14.4}{rm}$CCrr2: p_2(Z,f) \leftarrow r_1(X,Z)$}}}
\put(1951,-3436){\makebox(0,0)[lb]{\smash{\SetFigFont{12}{14.4}{rm}$\leftarrow r_1(X,Z), r_1(X,Z), p_2(Z,c)$}}}
\put(2551,-4036){\makebox(0,0)[lb]{\smash{\SetFigFont{12}{14.4}{rm}factor}}}
\put(4426,-5011){\makebox(0,0)[lb]{\smash{\SetFigFont{12}{14.4}{rm}$CCrr3: p_2(X,Y) \leftarrow r_2(X,Y)$}}}
\put(3601,-5461){\makebox(0,0)[lb]{\smash{\SetFigFont{12}{14.4}{rm}$\{X/Z, Y/c\}$}}}
\put(3526,-436){\makebox(0,0)[lb]{\smash{\SetFigFont{12}{14.4}{rm}$\{Y/W, Y'/c\}$}}}
\put(1951,-4636){\makebox(0,0)[lb]{\smash{\SetFigFont{12}{14.4}{rm}$\leftarrow r_1(X,Z), p_2(Z,c)$}}}
\put(1951,-5836){\makebox(0,0)[lb]{\smash{\SetFigFont{12}{14.4}{rm}$\leftarrow r_1(X,Z), r_2(Z,c)$}}}
\put(3601,-2986){\makebox(0,0)[lb]{\smash{\SetFigFont{12}{14.4}{rm}$\{X'/Z\}$}}}
\end{picture}
\caption{Functional Dependency That Cannot Be Handled by Decomposition}
\end{figure}

We note a generalization of the above result:\\
Suppose the {\bf EDB} consists of $n$ relations\\
$p_1(X,Y_2,\ldots,Y_n),p_2(X,Y_2),\ldots,p_n(X,Y_n)$ 
with the {\bf IC} containing the $n-1$ key constraints
$p_i: X \rightarrow Y_i$ 
for i = 2, $\ldots$, n  and the {\bf Res_{DB}} consisting of the
$n$ relations $r_1$, $r_2$, $\ldots$, $r_n$, defined as
\begin{equation}
r_1(X,Z) \leftarrow p_1(X,W_2,\ldots,W_n), p_2(Z,W_2), \ldots, p_n(Z,W_n)
\end{equation}
\begin{equation}
r_2(X,Y) \leftarrow p_2(X,Y) \ldots
\end{equation}
\hspace{3in} \vdots
\begin{equation}
r_n(X,Y) \leftarrow p_n(X,Y) \ldots
\end{equation}
where the $\ldots$ indicate the possibility of additional predicates.\\
Then the query
\begin{equation}
q(X): \leftarrow p_1(X,c_2,\ldots,c_n)
\end{equation}
can be folded as
\begin{equation}
\leftarrow r_1(X,Z), r_2(Z,c_2), \ldots, r_n(Z,c_n)
\end{equation}

\subsection{Functional Dependencies and Recursion}
The following example, discussed in \cite{DGL00:integration} illustrates how functional 
dependencies may introduce recursion.   We now consider how this is handled
in our approach and show that, although the non built-in predicates are
not recursive, recursion is introduced by the recursive transitivity rule 
of equality.

\begin{example}
{\bf EDB}: $schedule(Airline, Flight-No, Date, Pilot, AirCraft)$\\
{\bf IDB}: $\emptyset$\\
{\bf IC}: \\
\hspace*{.65in} $A_1 = A_2 \leftarrow s(A_1,N_1,D_1,P_1,C_1), s(A_2,N_2,D_2,P_1,C_2)$\\
\hspace*{.8in} (i.e., the functional dependency $Pilot \rightarrow Airline$)\\
\hspace*{.65in} $A_1 = A_2 \leftarrow s(A_1,N_1,D_1,P_1,C_1), s(A_2,N_2,D_2,P_2,C_1)$\\
\hspace*{.8in} (i.e., the functional dependency $Aircraft \rightarrow Airline$)\\
${\bf Res_{DB}}$:  $r(D,P,C) \leftarrow s(A,N,D,P,C)$\\
{\bf Query}: $q(P): \leftarrow s(A,N,D,mike,C), s(A,N',D',P,C')$\\
\\
$CCrr: s(f(D,P,C),g(D,P,C),D,P,C) \leftarrow r(D,P,C)$
\end{example}

In \cite{DGL00:integration}, they discuss how they obtain an infinite set of folded queries,
one for each $n$ of the form:\\\\
$q_n(P) \leftarrow r(D_1,mike,C_1), r(D_2,P_2,C_1), r(D_3,P_2,C_2), r(D_4,P_3,C_2), \ldots,$ \\
\hspace*{1.0in} $r(D_{2n-2},P_n,C_{n-1}), r(D_{2n-1},P_n,C_n), r(D_{2n}, P, C_n)$ \\\\

Using our approach we start with the expanded clause:\\\\
$\leftarrow s(A,N,D,P',C), s(A',N',D',P,C'), A = A', P' = mike$ \\\\
By applying the functional dependency $C \rightarrow A$, and factoring the
resolvent clause twice, and applying the $CCrr$ twice, we obtain the clause\\\\
$\leftarrow r(D,P',C), r(D',P,C), P' = mike$\\\\
which is equivalent to $q_1(P), i>1$.  Although we do not provide the details here, 
in order to obtain the other $q_i(P)$, we need to use the recursive transitivity rule of equality
as well as the two integrity constraints several times and factoring several times..

\subsection{Inclusion Dependency Example}

The last example of this section illustrates the use of an inclusion
dependency.
This example is taken from Example 4.3 of \cite{Gryz98}
and contains an example given earlier with some modifications:
The {\bf EDB} contains four relations:\\
{\bf Patient} (name,dob,address,insurer)\\
{\bf Procedure} (patient_name,physician_name,procedure_name,time)\\
{\bf Insurer} (company,address,phone)\\
{\bf Event} (event_name,description,patient_name,location)\\
The {\bf IDB} is empty.\\
The {\bf IC} contains a single inclusion dependency:\\
{\em Procedure(procedure_name,patient_name)} $\subseteq$ 
{\em Event(event_name,patient_name)}\\
written in {\bf Datalog} as\\
$event(X'_3,f_1,X'_1,g) \leftarrow procedure(X'_1,X'_2,X'_3,X'_4)$.\\
As before, the {\bf Res_{DB}} consists of two relations
{\bf Clinical_History} and {\bf Billing}, written as
$r_1$ and $r_2$ with the following definitions:
\begin{equation}
r_1(X_1,X_2,X_3,X_4) \leftarrow patient(X_1,X_2,U_1,U_2), 
procedure(X_1,U_3,X_3,X_4)
\end{equation}
\begin{equation}
r_2(Y_1,Y_2,Y_3) \leftarrow patient(Y_1,V_1,Y_2,V_2), insurer(V_2,Y_3,V_3)
\end{equation}
Preprocessing yields the same Clark completion resource rules as before:
\begin{equation}
CCrr1:  patient(X_1,X_2,f_1,f_2) \leftarrow r_1(X_1,X_2,X_3,X_4)
\end{equation}
\begin{equation}
CCrr2:  procedure(X_1,f_3,X_3,X_4) \leftarrow r_1(X_1,X_2,X_3,X_4)
\end{equation}
\begin{equation}
CCrr3:  patient(Y_1,g_1,Y_2,g_2) \leftarrow r_2(Y_1,Y_2,Y_3)
\end{equation}
\begin{equation}
CCrr4:  insurer(g_2,Y_3,g_3) \leftarrow r_2(Y_1,Y_2,Y_3)
\end{equation}
The query asks for the names of events recorded for patients born
before 1930:
\begin{equation}
q(X_3): \leftarrow patient(X_1,X_2,Y_1,Y_2), event(X_3,Y_3,X_1,Y_4), X_2 \leq
1930
\end{equation}
The derivation is shown in 
Figure 7.
Using the inclusion dependency and two Clark completion resource rules
we obtain the answer as
\begin{equation}
\leftarrow r_1(X_1,X_2,X_3,X_4), X_2 \leq 1930
\end{equation}
\begin{figure}
\label{inclusion}
\setlength{\unitlength}{0.00083300in}%
\begingroup\makeatletter\ifx\SetFigFont\undefined
\def\x#1#2#3#4#5#6#7\relax{\def\x{#1#2#3#4#5#6}}%
\expandafter\x\fmtname xxxxxx\relax \def\y{splain}%
\ifx\x\y   
\gdef\SetFigFont#1#2#3{%
  \ifnum #1<17\tiny\else \ifnum #1<20\small\else
  \ifnum #1<24\normalsize\else \ifnum #1<29\large\else
  \ifnum #1<34\Large\else \ifnum #1<41\LARGE\else
     \huge\fi\fi\fi\fi\fi\fi
  \csname #3\endcsname}%
\else
\gdef\SetFigFont#1#2#3{\begingroup
  \count@#1\relax \ifnum 25<\count@\count@25\fi
  \def\x{\endgroup\@setsize\SetFigFont{#2pt}}%
  \expandafter\x
    \csname \romannumeral\the\count@ pt\expandafter\endcsname
    \csname @\romannumeral\the\count@ pt\endcsname
  \csname #3\endcsname}%
\fi
\fi\endgroup
\begin{picture}(3150,5745)(1201,-5125)
\thicklines
\put(2401,389){\line( 0,-1){ 75}}
\put(2401,314){\line( 0,-1){975}}
\put(2401,-586){\line( 0, 1){ 75}}
\put(2401,-661){\line( 3, 1){1867.500}}
\put(2401,-1036){\line( 0,-1){ 75}}
\put(2401,-1111){\line( 0,-1){975}}
\put(2401,-2075){\line( 3, 1){1867.500}}
\put(2401,-2461){\line( 0,-1){ 75}}
\put(2401,-2536){\line( 0,-1){975}}
\put(2401,-3477){\line( 3, 1){1867.500}}
\put(2401,-2836){\makebox(6.6667,10.0000){\SetFigFont{10}{12}{rm}.}}
\put(2401,-3811){\line( 0,-1){ 75}}
\put(2401,-3886){\line( 0,-1){975}}
\put(1201,464){\makebox(0,0)[lb]{\smash{\SetFigFont{12}{14.4}{rm}$Query: \leftarrow patient(X_1,X_2,Y_1,Y_2), event(X_3,Y_3,X_1,Y_4), X_2 \leq 1930$}}}
\put(4351, 14){\makebox(0,0)[lb]{\smash{\SetFigFont{12}{14.4}{rm}$IC: event(X'_3,f,X'_1,g) \leftarrow procedure(X'_1,X'_2,X'_3,X'_4)$}}}
\put(1201,-961){\makebox(0,0)[lb]{\smash{\SetFigFont{12}{14.4}{rm}$\leftarrow patient(X_1,X_2,Y_1,Y_2), procedure(X_1,X"_2,X_3,X'_4), X_2 \leq 1930$}}}
\put(3601,-361){\makebox(0,0)[lb]{\smash{\SetFigFont{12}{14.4}{rm}$\{X'_3/X_3, Y_3/f, X'_1/X_1, Y_4/g\}$}}}
\put(4351,-1411){\makebox(0,0)[lb]{\smash{\SetFigFont{12}{14.4}{rm}$CCrr1: patient(X_1,X_2,f_1,f_2) \leftarrow r_1(X_1,X_2,X_3,X_4)$}}}
\put(3601,-1861){\makebox(0,0)[lb]{\smash{\SetFigFont{12}{14.4}{rm}$\{Y_1/f_1, Y_2/f_2\}$}}}
\put(1201,-2386){\makebox(0,0)[lb]{\smash{\SetFigFont{12}{14.4}{rm}$\leftarrow r_1(X_1,X_2,X_3,X_4), procedure(X_1,X'_2,X_3,X'_4), X_2 \leq 1930$ }}}
\put(4351,-2836){\makebox(0,0)[lb]{\smash{\SetFigFont{12}{14.4}{rm}$CCrr2: procedure(X_1,f_3,X_3,X_4) \leftarrow r_1(X_1,X_2,X_3,X_4)$}}}
\put(3601,-3286){\makebox(0,0)[lb]{\smash{\SetFigFont{12}{14.4}{rm}$\{X'_2,f_3,X'_4/X_4\}$}}}
\put(1201,-3736){\makebox(0,0)[lb]{\smash{\SetFigFont{12}{14.4}{rm}$\leftarrow r_1(X_1,X_2,X_3,X_4), r_1(X_1,X_2,X_3,X_4), X_2 \leq 1930$}}}
\put(2551,-4261){\makebox(0,0)[lb]{\smash{\SetFigFont{12}{14.4}{rm}factor}}}
\put(1201,-5086){\makebox(0,0)[lb]{\smash{\SetFigFont{12}{14.4}{rm}$\leftarrow r_1(X_1,X_2,X_3,X_4), X_2 \leq 1930$}}}
\end{picture}
\caption{Inclusion Dependency Example}
\end{figure}
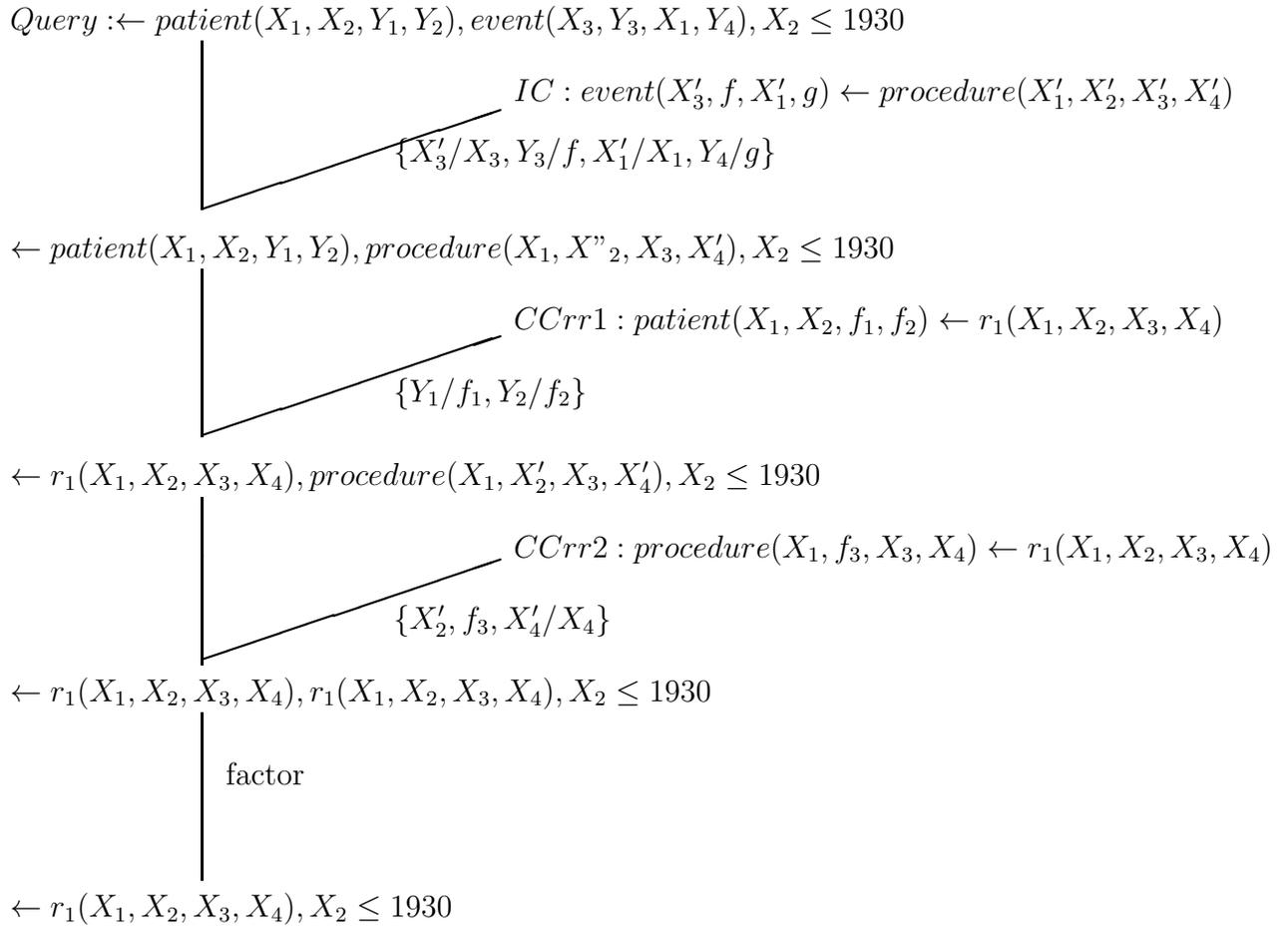

\section{Multiple Definitions for Resources}\label{Sec:mult-res}
In this section we consider the case where there are resources that
were developed from several definitions or queries.  
This is in contrast to work in the previous section where it was assumed
that a resource was constructed from a single conjunctive view.  
We also allow queries that are in disjunctive normal form, that is,
the queries involve disjunctions of conjunctions.
Our assumptions for the database are different from the ones given
at the beginning of Section~\ref{Sec:fold}.  We list our assumptions here.

\begin{enumerate}
  \item No formula contains negation.
  \item Each {\bf IDB} predicate may be defined by multiple safe,
conjunctive, non-recursive function-free Horn rules.
  \item Each ${\bf Res_{DB}}$ predicate is defined by a set of
safe conjunctive function-free Horn formulas on {\bf EDB} and/or {\bf IDB} predicates.
  \item Each {\bf IC} clause is a safe function-free formula of the
form $~~~G \leftarrow F~~~$
where $G$ is a disjunction of zero or more {\bf EDB} predicates 
and $F$ is a conjunction of {\bf EDB} predicates.
  \item Each query has the form $~~~q: \leftarrow G~~~$ where $G$ is 
in disjunctive normal form on {\bf EDB} and {\bf IDB} predicates.
  \item The database includes axioms for built-in predicates as needed.
\end{enumerate}

Resource definitions may contain constants.  We assume that the resource
rules are expanded, and rectified (see page~\pageref{expanded})
so that a single resource that has multiple definitions
is defined by the same variables in each definition.

We first provide a simple 
example, that illustrates what has to be done to handle such cases.

\begin{example}[Multi-Resource Example]\label{mult-resource}
{\bf EDB}: $p_1(X)$, $p_2(X)$, $p_3(X).$\\ 
{\bf IDB}: $\emptyset$.\\
\begin{equation}
{\bf IC}:  p_2(X) \lor p_3(X) \Leftarrow p_1(X). \label{eq:r1} \\
\end{equation}
The integrity constraint is a non-Horn clause and states that whenever
$p_1(a)$ is in the database, for some constant $a$, then either
$p_2(a)$ or $p_3(a)$ or both are in the database.  We also have the
following resource definitions:\\
${\bf Res_{DB}}$:\\
\begin{equation}
r(X) \gets p_1(X), p_2(X).\label{eq:r16}\\
\end{equation}
\begin{equation}
r(X) \gets p_3(X).\label{eq:r3}\\
\end{equation}
The Clark completion of the resource rules (ignoring equality) is:\\
\begin{equation}
r(X) \leftrightarrow (p_1(X) \land p_2(X)) \lor p_3(X)
\end{equation}
After the use of De Morgan's rules we obtain two Clark completion
resource rules:\\
\begin{equation}
CCrr1: p_1(X) \lor p_3(X) \gets r(X). \label{eq:r5},
\end{equation}
and
\begin{equation}
CCrr2: p_2(X) \lor p_3(X) \gets r(X). \label{eq:r6}
\end{equation}
\end{example}

Note that the use of more than one conjunctive definition of a resource
leads to non-Horn clauses.  Such clauses are outside of {\bf Datalog}.
Hence to work with such clauses, we need, in general {\bf Disjunctive Datalog},
that is {\bf $Datalog^{\lor}$}.  

Next, consider the query:
\begin{equation}
q(X): \gets p_1(X) \lor p_3(X). \label{eq:r4}\\
\end{equation}    

\begin{figure}
\label{resource-deriv}
\setlength{\unitlength}{0.00083300in}%
\begingroup\makeatletter\ifx\SetFigFont\undefined
\def\x#1#2#3#4#5#6#7\relax{\def\x{#1#2#3#4#5#6}}%
\expandafter\x\fmtname xxxxxx\relax \def\y{splain}%
\ifx\x\y   
\gdef\SetFigFont#1#2#3{%
  \ifnum #1<17\tiny\else \ifnum #1<20\small\else
  \ifnum #1<24\normalsize\else \ifnum #1<29\large\else
  \ifnum #1<34\Large\else \ifnum #1<41\LARGE\else
     \huge\fi\fi\fi\fi\fi\fi
  \csname #3\endcsname}%
\else
\gdef\SetFigFont#1#2#3{\begingroup
  \count@#1\relax \ifnum 25<\count@\count@25\fi
  \def\x{\endgroup\@setsize\SetFigFont{#2pt}}%
  \expandafter\x
    \csname \romannumeral\the\count@ pt\expandafter\endcsname
    \csname @\romannumeral\the\count@ pt\endcsname
  \csname #3\endcsname}%
\fi
\fi\endgroup
\begin{picture}(6612,4020)(3001,-3325)
\thicklines
\put(6601,464){\line(-5,-1){3000}}
\put(3601,-136){\line( 1, 0){ 75}}
\put(6601,464){\line( 5,-1){3000}}
\put(3676,-1561){\line( 5, 2){1875}}
\put(3676,-286){\line( 0,-1){1275}}
\put(3676,-1786){\line( 0,-1){1275}}
\put(3676,-3061){\line( 5, 2){6012.931}}
\put(4801,539){\makebox(0,0)[lb]{\smash{\SetFigFont{12}{14.4}{rm}$Query: \leftarrow p_1(X) \vee p_3(X)$}}}
\put(3001,-286){\makebox(0,0)[lb]{\smash{\SetFigFont{12}{14.4}{rm}$\leftarrow p_1(X)$}}}
\put(9001,-361){\makebox(0,0)[lb]{\smash{\SetFigFont{12}{14.4}{rm}$\leftarrow p_3(X)$}}}
\put(3001,-1786){\makebox(0,0)[lb]{\smash{\SetFigFont{12}{14.4}{rm}$p_3(X) \leftarrow r(X)$}}}
\put(3001,-3286){\makebox(0,0)[lb]{\smash{\SetFigFont{12}{14.4}{rm}$\leftarrow r(X)$}}}
\put(5176,-736){\makebox(0,0)[lb]{\smash{\SetFigFont{12}{14.4}{rm}$CCrr1: p_1(X) \vee p_3(X) \leftarrow r(X)$}}}
\end{picture}
\caption{Deriving Resource Queries}
\end{figure}
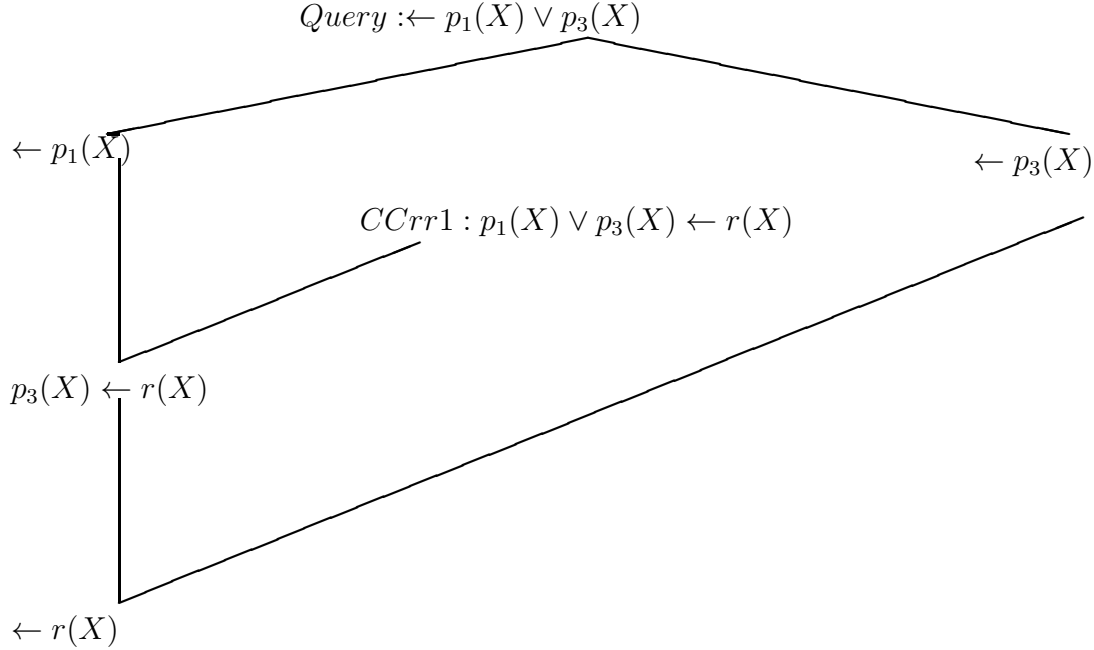

Now we have to determine if we can derive
answers to the query from the {\bf EDB}, the {\bf IDB}, and the 
{\bf CCrr} (that is, rule~\ref{eq:r5} and 
rule~\ref{eq:r6}).
Figure 8
shows the steps required to achieve this result.
Because the query is a disjunction,
$\gets p_1(X) \lor p_3(X)$,
the derivation can be split into two clauses:
$\gets p_1(X)$ and $\gets p_3(X)$.
The derivation terminates with a clause that contains only 
the resource predicate.  
We will show that all answers obtained by querying the resource 
will yield correct answers to the query $q(X)$. 

\noindent
\begin{example}[Example~\ref{mult-resource} Continued]

Consider the same example, where the query was:
\\
\begin{equation}
q(X): \leftarrow p_1(X).
\end{equation}

\noindent
Figure 8 applies without the right branch $\leftarrow p_3(X)$.
We obtain as bottom clause,\\
$p_3(X) \leftarrow r(X)$.  
If written with an empty head, we
obtain $\leftarrow r(X), \neg p_3(X)$.  
This is a
query that contains logical negation.  We cannot replace this 
by default negation since default negation does not imply
logical negation.  We can see the problem by assuming that
we have $r(c), p_3(a), p_3(b)$.  $p_3(c)$ may either be {\em true}
or {\em false}, but default negation assumes its falsity.
We might  have $r(c), \neg p_3(c)$ is not {\em true}, while $r(c), not~p_3(c)$
is {\em true}.  Hence we could obtain a non-sound answer $q(c)$
by using default negation.
However, if one knew that the $p_3$ predicate were complete
(by the Closed World Assumption) one could write an axiom
\begin{equation}
\neg p_3 \leftarrow not~p_3 
\end{equation}
in which case, another resolution step would be applied to obtain
\begin{equation}
\leftarrow r(X), not~p_3(X)
\end{equation}
\end{example}

The above query is a partial folding.  The negation of the atom $p_3(X)$
does not lead to complications since the resource $r(X)$ makes the formula
safe.  

The folding algorithm we now present uses set-of-support as its inference method.  In this
inference, there are two types of clauses: one set, referred to as
$T$, is given support.  This set consists of the conjunctions that
are part of the query.  The second set consists of a satisfiable set of
clauses, the remaining clauses in the set $\cal C$, which we may refer to
as set $S$.  A set of support resolution is a resolution of two clauses 
that are not both from $S - T$ (\cite{Chan73}).

Before the algorithm commences we assume that there is a test to determine
which of the three cases (see page 10) applies to {\bf ICs}.  In Case 1, 
nothing has to be done.  In Case 2,
we assume that the set-of-support derivation is modified to include a check to determine
if the clause, $L$, that has been generated, satisfies
the bounding condition, and if it does, then backtracking occurs.  In Case 3, a
depth bound $k$ is specified and if the depth is reached, backtracking occurs.
We also assume that if there are built-in predicates such as $=, \not=, \ge$,
then the input clauses are placed in expanded form.  If there are no built-in
predicates, it is unnecessary to do the expansion.

\begin{description}
\item[Folding Algorithm 2 (Finding Multiple Foldings)]\label{folding2}
\item[Input:] $\cal C$: the set of clauses in the {\bf EDB}, {\bf IDB},
{\bf IC}, and {\bf CCrr}, and\\
the query, $q(\bar{X})$: $\leftarrow G(\bar{Y})$, where $\bar{X} \subseteq \bar{Y}$,
and $G$ is in disjunctive normal form.
\item[Output:] A proof tree starting with the query.
\item {\bf begin}\\
Split the query, $\leftarrow G$, into a set of clauses, (each of which has support).
Find a proof tree using set-of-support resolution
that results in leaf nodes of the form $\leftarrow L$ that contains 
the query variables and no function symbols. 
\item {\bf end}
\end{description}

We consider four cases for the clauses that are leaf nodes in the proof tree.
\begin{enumerate}
\item The clause has an empty head and a conjunction of resource and built-in
predicates in the body.  
\item The clause has an empty head and a conjunction of resource, {\bf EDB} and built-in
predicates in the body.  
\item The clause has no resource predicates.
\item The clause has a non-empty head.
\end{enumerate}

\begin{theorem}
Consider a clause that is a leaf node of the proof tree constructed by 
Folding Algorithm 2.  In Cases 1 and 2, the query can be answered using the
clause and every answer obtained that way is sound.  These cases are instances
of complete and partial folding respectively.  In Case 3, the clause is not a
folding.  In Case 4, the clause is not a folding, but if the $CWA$ can be applied
to all the predicates in the head, then the atoms in the head can be moved to the
body using default negation and Case 2 or Case 3 becomes applicable.
\end{theorem}

\begin{proof}
For a clause of Case 1 or 2, the proof of Theorem~\ref{sound} applies.
In Case 3, there is no folding.  In Case 4, without the $CWA$, the clause
is not a query (since it does not have an empty head).  The disjunction of
atoms in the head of such a clause can be moved to the body 
and negated (for example, by $\neg p$).  Since the $CWA$ applies, the
axioms  $\neg p \leftarrow not~p$ can be added to the set of clauses $\cal C$.
These axioms may be used to eliminate the logically negated atoms and 
replaced by default negated atoms.  When this is done, this case reduces to
Case 2 or Case 3.
\end{proof}

We illustrate the theorem with an example:

\begin{example}
Consider the {\bf EDB} with the relations $p_1(X_1,Y_1), \ldots, p_7(X_7,Y_7)$
Assume there are no {\bf IDB} predicates and no {\bf ICs}.  Let there be the following 
resource rules:
\\
$r_1(X) \leftarrow p_1(X,Z), p_2(X,Z), Z \neq a$
\\
$r_2(X,Y) \leftarrow p_5(X,Z), p_6(Z,Y)$
\\
$r_2(X,Y) \leftarrow p_7(X,Y)$

The Clark Completion resource rules are:
\begin{equation}
CCrr1: p_1(X,f) \leftarrow r_1(X)
\end{equation}
\begin{equation}
CCrr2: p_2(X,f) \leftarrow r_1(X)
\end{equation}
\begin{equation}
CCrr3: f \neq a \leftarrow r_1(X)
\end{equation}
\begin{equation}
CCrr4: p_5(X,f) \vee p_7(X,Y) \leftarrow r_2(X,Y)
\end{equation}
\begin{equation}
CCrr5:p_6(X,f) \vee p_7(X,Y) \leftarrow r_2(X,Y)
\end{equation}

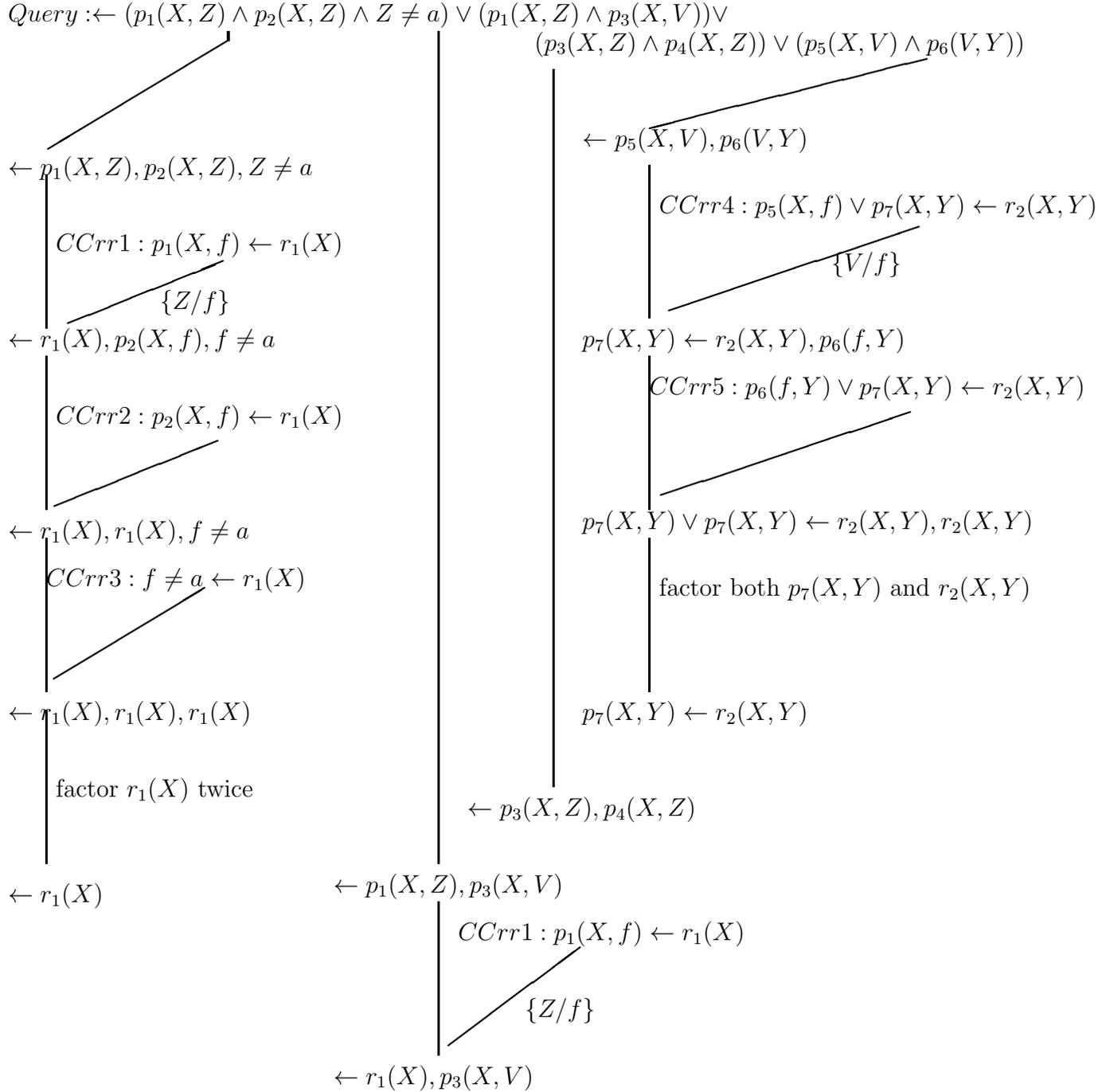
\begin{figure}
\label{theorem-example}
\setlength{\unitlength}{0.00083300in}%
\begingroup\makeatletter\ifx\SetFigFont\undefined
\def\x#1#2#3#4#5#6#7\relax{\def\x{#1#2#3#4#5#6}}%
\expandafter\x\fmtname xxxxxx\relax \def\y{splain}%
\ifx\x\y   
\gdef\SetFigFont#1#2#3{%
  \ifnum #1<17\tiny\else \ifnum #1<20\small\else
  \ifnum #1<24\normalsize\else \ifnum #1<29\large\else
  \ifnum #1<34\Large\else \ifnum #1<41\LARGE\else
     \huge\fi\fi\fi\fi\fi\fi
  \csname #3\endcsname}%
\else
\gdef\SetFigFont#1#2#3{\begingroup
  \count@#1\relax \ifnum 25<\count@\count@25\fi
  \def\x{\endgroup\@setsize\SetFigFont{#2pt}}%
  \expandafter\x
    \csname \romannumeral\the\count@ pt\expandafter\endcsname
    \csname @\romannumeral\the\count@ pt\endcsname
  \csname #3\endcsname}%
\fi
\fi\endgroup
\begin{picture}(7210,8520)(76,-7750)
\thicklines
\put(1801,464){\line( 0, 1){ 75}}
\put(1801,464){\line(-5,-3){1411.765}}
\put(376,-586){\line( 0,-1){1200}}
\put(544,-1750){\line( 5, 2){1215.517}}
\put(376,-2011){\line( 0,-1){1200}}
\put(439,-3181){\line( 5, 2){1280.172}}
\put(376,-3436){\line( 0,-1){1200}}
\put(435,-4533){\line( 5, 3){1180.147}}
\put(376,-4786){\line( 0,-1){1200}}
\put(3451,539){\line( 0,-1){6525}}
\put(3451,-6286){\line( 0,-1){1200}}
\put(3526,-7411){\line( 4, 3){1032}}
\put(4351,239){\line( 0,-1){5625}}
\put(5101,-511){\line( 0,-1){1200}}
\put(5101,-2011){\line( 0,-1){1200}}
\put(5101,-3436){\line( 0,-1){1200}}
\put(5255,-1646){\line( 3, 1){1957.500}}
\put(5189,-3097){\line( 3, 1){1957.500}}
\put(7274,324){\line(-4,-1){2170.588}}
\put( 76,614){\makebox(0,0)[lb]{\smash{\SetFigFont{12}{14.4}{rm}$Query: \leftarrow (p_1(X,Z) \wedge p_2(X,Z) \wedge Z \neq a) \vee (p_1(X,Z) \wedge p_3(X,V)) \vee$}}}
\put( 76,-586){\makebox(0,0)[lb]{\smash{\SetFigFont{12}{14.4}{rm}$\leftarrow p_1(X,Z), p_2(X,Z), Z\neq a$}}}
\put(451,-1186){\makebox(0,0)[lb]{\smash{\SetFigFont{12}{14.4}{rm}$CCrr1: p_1(X,f) \leftarrow r_1(X)$}}}
\put(1201,-1636){\makebox(0,0)[lb]{\smash{\SetFigFont{12}{14.4}{rm} $\{Z/f\}$}}}
\put(451,-2536){\makebox(0,0)[lb]{\smash{\SetFigFont{12}{14.4}{rm}$CCrr2: p_2(X,f) \leftarrow r_1 (X)$}}}
\put( 76,-1936){\makebox(0,0)[lb]{\smash{\SetFigFont{12}{14.4}{rm}$\leftarrow r_1(X), p_2(X,f), f \neq a$}}}
\put( 76,-3436){\makebox(0,0)[lb]{\smash{\SetFigFont{12}{14.4}{rm}$\leftarrow r_1(X), r_1(X), f \neq a$}}}
\put(376,-3811){\makebox(0,0)[lb]{\smash{\SetFigFont{12}{14.4}{rm}$CCrr3: f \neq a \leftarrow r_1(X)$}}}
\put( 76,-4861){\makebox(0,0)[lb]{\smash{\SetFigFont{12}{14.4}{rm}$\leftarrow r_1(X), r_1(X), r_1(X)$}}}
\put( 76,-6286){\makebox(0,0)[lb]{\smash{\SetFigFont{12}{14.4}{rm}$\leftarrow r_1(X)$}}}
\put(2626,-6211){\makebox(0,0)[lb]{\smash{\SetFigFont{12}{14.4}{rm}$\leftarrow p_1(X,Z), p_3(X,V)$}}}
\put(3601,-6586){\makebox(0,0)[lb]{\smash{\SetFigFont{12}{14.4}{rm}$CCrr1: p_1(X,f) \leftarrow r_1(X)$}}}
\put(2626,-7711){\makebox(0,0)[lb]{\smash{\SetFigFont{12}{14.4}{rm}$\leftarrow r_1(X), p_3(X,V)$}}}
\put(4126,-7186){\makebox(0,0)[lb]{\smash{\SetFigFont{12}{14.4}{rm}$\{Z/f\}$}}}
\put(4201,389){\makebox(0,0)[lb]{\smash{\SetFigFont{12}{14.4}{rm}$(p_3(X,Z) \wedge p_4(X,Z)) \vee (p_5(X,V) \wedge p_6(V,Y))$}}}
\put(3676,-5611){\makebox(0,0)[lb]{\smash{\SetFigFont{12}{14.4}{rm}$\leftarrow p_3(X,Z), p_4(X,Z)$}}}
\put(4576,-1936){\makebox(0,0)[lb]{\smash{\SetFigFont{12}{14.4}{rm}$p_7(X,Y) \leftarrow r_2(X,Y), p_6(f,Y)$}}}
\put(4576,-361){\makebox(0,0)[lb]{\smash{\SetFigFont{12}{14.4}{rm}$\leftarrow p_5(X,V), p_6(V,Y)$}}}
\put(4576,-3361){\makebox(0,0)[lb]{\smash{\SetFigFont{12}{14.4}{rm}$p_7(X,Y) \vee p_7(X,Y) \leftarrow r_2(X,Y), r_2(X,Y)$}}}
\put(5176,-886){\makebox(0,0)[lb]{\smash{\SetFigFont{12}{14.4}{rm}$CCrr4: p_5(X,f) \vee p_7(X,Y) \leftarrow r_2(X,Y)$}}}
\put(6526,-1336){\makebox(0,0)[lb]{\smash{\SetFigFont{12}{14.4}{rm}$\{V/f\}$}}}
\put(5101,-2311){\makebox(0,0)[lb]{\smash{\SetFigFont{12}{14.4}{rm}$CCrr5: p_6(f,Y) \vee p_7(X,Y) \leftarrow r_2(X,Y)$}}}
\put(4576,-4861){\makebox(0,0)[lb]{\smash{\SetFigFont{12}{14.4}{rm}$p_7(X,Y) \leftarrow r_2(X,Y)$}}}
\put(451,-5461){\makebox(0,0)[lb]{\smash{\SetFigFont{12}{14.4}{rm}factor $r_1(X)$ twice}}}
\put(5176,-3886){\makebox(0,0)[lb]{\smash{\SetFigFont{12}{14.4}{rm}factor both $p_7(X,Y)$ and $r_2(X,Y)$}}}
\end{picture}
\caption{Four Cases of Theorem~\ref{sound} Illustrated}
\end{figure}

Let the query be given by:
\\
$q(X) \leftarrow p_1(X,Z) \wedge ((p_2(X,Z) \wedge Z \neq a) \vee p_3(X,V)) 
\vee (p_3(X,Z) \wedge p_4(X,Z)) \vee (p_5(X,V) \wedge p_6(V,Y))$
We construct the proof tree in 
Figure 9 after converting the
query to disjunctive normal form.
There are four leaf nodes in 
Figure 9. The leftmost leaf node
has only resource predicates, and hence, it can be used to obtain correct answers.
The second leftmost leaf node has both a resource predicate and an extensional predicate.
Hence, it is a partial folding and if the database and the resource predicate were used,
correct answers would be found.  The third leftmost leaf provides nothing with respect 
to resources. 
The final leaf node has something in the head of the clause and 
provides no useful information (without the Closed World Assumption
on $p_7$).  
If the CWA applies to $p_7$ then we obtain a partial folding:\\
$\leftarrow r_2(X,Y), not~p_7(X,Y)$.
\end{example}


In the case of the query $q(X,Y,Z,W)$ given at the end of Section 4.1,
no resolution steps are possible, hence Case 3 applies.

Answering the query with resource rules as described in the above theorem
does not mean that all answers to the original query have been found.  
We want to determine when using resource predicates (or partially using resource
predicates) will provide all the answers, that is, 
if the method is complete.  This assumes that the database from which the 
resources were constructed has not been updated.

When we do not have completeness using resource predicates we may still wish to
find all answers.  This may be done as follows.  Let the answers be given by the
formula defining $Q_{resource}$, and let the query be $Q$.  Then the remaining 
answers may be found by using the query: $Q - Q_{resource}$.  In some cases this
may be simpler than trying to find all answers to $Q$.  For example, if the
query $Q$ is given by $Q: (p_4 \wedge p_5) \vee (p_1 \wedge p_2) \vee (p_2 \wedge p_3)$,
and $r_1: p_1 \wedge p_2$ and $r_2: p_2 \wedge p_3$.
then the remaining answers can be expressed by $p_4 \wedge p_5$, and hence it is
easier in this case to find the complete set of answers to the query by using this subformula
together with the answers found by $Q_{resource}$ 
than having to answer the original query, $Q$.

In the following algorithm, where we test for completeness, we have to consider
two cases.  In the first case there are no built-in predicates.   In the 
second case, we allow built-in predicates.  In this case we have a bound
on the depth of the proof tree.  This algorithm is based on the subsumption algorithm.
In the subsumption algorithm two clauses, $C$ and $D$ are given and the algorithm
checks to determine if $C$ subsumes $D$, which means that there is a substitution
$\theta$ such that $C\theta$ is a subset of $D$, where the clauses $C$ and $D$ are
considered as sets of literals.  In our completeness test, $D$ corresponds to the original query
and the $C$ is the union of the leaf nodes.  What we are trying to show is that the
original query is implied by the union of the leaf nodes using the
{\bf IDB}, {\bf IC}, and ${\bf Res_{DB}}$.

\begin{description}
\item [Completeness Test Algorithm]\label{completeness-test}
\item {\bf Input:} The {\bf IDB}, {\bf IC}, ${\bf Res_{DB}}$, the non-negated form of
the query, where each variable is replaced by a unique new constant, and the leaf nodes 
of Cases 1 and 2 of Folding Algorithm 2.
If there are built-in predicates, the appropriate axioms, such as equality, 
are also included as well as a bound on the depth of the proof tree.
\item {\bf begin} \\
We give support to the clauses of the leaf nodes obtained in Algorithm~\ref{folding2},
and apply set-of support resolution on the input clauses.  If we obtain the
null clause we terminate.  If we allow built-in predicates, when we reach 
the depth bound, we terminate.
\item {\bf end}
\end{description}

In the following theorem we show when completeness is obtained.


\begin{theorem}
\label{complete}
If the Completeness Test Algorithm yields the null
clause, then the set of answers obtained by solving the folded queries is
complete.  If the Completeness Test Algorithm ends without reaching the 
null clause, then the set of answers obtained by solving the folded queries
may not be complete.
\end{theorem}


\begin{proof}
Suppose that the null clause has been found.  Let $\bar{a}$ be a solution
to $q(\bar{X})$.  Writing $q(\bar{X}): \leftarrow G(\bar{Y})$, we obtain 
$DB \models G(\bar{b})$, where $\bar{b}[\bar{X}] = \bar{a}$.  
That is, the constants obtained by answering the query only return that 
part of the tuple required by the variables in $\bar{X}$.  The projection of
$\bar{b}$ on the variables in $\bar{X}$ yields $\bar{a}$.
Suppose that the leaf nodes used in the completeness test are 
$\leftarrow L_{1}(\bar{Z_{1}}), \ldots, \leftarrow L_{n}(\bar{Z_{n}})$,
where $\bar{X} \subset \bar{Z_{i}}$ for $1 \leq i \leq n$. Thus,
\begin{center}
{\bf IDB} $\cup$ {\bf IC} $\cup$ ${\bf Res_{DB}}$ $\cup$ $\{G(\bar{b})\} \cup
\leftarrow L_{1}(\bar{Z_{1}}) \cup \ldots \cup \leftarrow L_{n}(\bar{Z_{n}})$
\end{center}
form a contradiction.  Hence, 
\begin{center}
$DB \models$ ({\bf IDB} $\cup$ {\bf IC} $\cup$ ${\bf Res_{DB}}$ $\cup$ $\{G(\bar{b})\})
\rightarrow L_{1}(\bar{Z_{1}}) \vee \ldots \vee L_{n}(\bar{Z_{n}})$.
\end{center}
Since $DB \models$ {\bf IDB} $\cup$ {\bf IC} $\cup$ ${\bf Res_{DB}}$ $\cup$ $\{G(\bar{b})\}$,\\
$DB \models L_{1}(\bar{d_{1}}) \vee \ldots \vee L_{n}(\bar{d_{n}})$, for some
$\bar{d_{i}}$, $1 \leq i \leq n$, where $\bar{d_{i}}[\bar{X}] = \bar{a}$. Hence, the solution 
$\bar{a}$ to $q(\bar{X})$ is also obtained by one of the $L_{i}, 1 \leq i \leq n$.

Suppose the null clause is not obtained.  Proceeding as in the previous case, given
a solution $\bar{a}$ to $q(\bar{X})$ we cannot prove that for some 
$\bar{d_{i}}$, where $\bar{d_{i}}[\bar{X}] = \bar{a}$, 
$DB \models L_{1}(\bar{d_{1}}) \vee \ldots \vee L_{n}(\bar{d_{n}})$.
Hence, it is possible that the solution $\bar{a}$ to $q(\bar{X})$ may not
be obtained by solving the folded query.
\end{proof}

Note that if the null clause is found in Theorem~\ref{complete}, and all of 
the leaf nodes are of case 1, we have query completeness for a complete folding, 
otherwise we have query completeness for a partial folding.

We illustrate this theorem by reconsidering the example at the beginning
of this section.
The derivation is shown in Figure 10.  The original query, $\leftarrow p_1(X) \vee p_3(X)$
is changed to its non-negated form with the new constant $k$ substituted for the
variable $X$ to become $p_1(k) \vee p_3(k) \leftarrow$.
We start with the rewritten query, the leaf node, $\leftarrow r(X)$, and eventually
obtain the null clause. This shows that the query using the resource
predicate obtains all the answers to the original query.
Both factoring and ancestry-resolution are used.

\begin{figure}
\label{ancestry}
\setlength{\unitlength}{0.00083300in}%
\begingroup\makeatletter\ifx\SetFigFont\undefined
\def\x#1#2#3#4#5#6#7\relax{\def\x{#1#2#3#4#5#6}}%
\expandafter\x\fmtname xxxxxx\relax \def\y{splain}%
\ifx\x\y   
\gdef\SetFigFont#1#2#3{%
  \ifnum #1<17\tiny\else \ifnum #1<20\small\else
  \ifnum #1<24\normalsize\else \ifnum #1<29\large\else
  \ifnum #1<34\Large\else \ifnum #1<41\LARGE\else
     \huge\fi\fi\fi\fi\fi\fi
  \csname #3\endcsname}%
\else
\gdef\SetFigFont#1#2#3{\begingroup
  \count@#1\relax \ifnum 25<\count@\count@25\fi
  \def\x{\endgroup\@setsize\SetFigFont{#2pt}}%
  \expandafter\x
    \csname \romannumeral\the\count@ pt\expandafter\endcsname
    \csname @\romannumeral\the\count@ pt\endcsname
  \csname #3\endcsname}%
\fi
\fi\endgroup
\begin{picture}(2700,7827)(4126,-7144)
\thicklines
\put(5401,539){\line( 0,-1){675}}
\put(5401,-136){\line( 3, 1){1282.500}}
\put(5401,-361){\line( 0,-1){675}}
\put(5462,-997){\line( 3, 1){1282.500}}
\put(5401,-1261){\line( 0,-1){675}}
\put(5401,-2161){\line( 0,-1){675}}
\put(5401,-2161){\line( 0,-1){675}}
\put(5462,-2805){\line( 3, 1){1282.500}}
\put(5401,-3211){\line( 0,-1){675}}
\put(5401,-4261){\line( 0,-1){675}}
\put(5401,-5311){\line( 0,-1){675}}
\put(5462,-4882){\line( 3, 1){1282.500}}
\put(5462,-5981){\line( 3, 1){1282.500}}
\put(5401,-6286){\line( 0,-1){675}}
\put(4801,539){\makebox(0,0)[lb]{\smash{\SetFigFont{12}{14.4}{rm}$\leftarrow r(X)$}}}
\put(4801,-286){\makebox(0,0)[lb]{\smash{\SetFigFont{12}{14.4}{rm}$\leftarrow p_1(X), p_2(X)$}}}
\put(5701,-1561){\makebox(0,0)[lb]{\smash{\SetFigFont{12}{14.4}{rm}factor}}}
\put(4126,-1186){\makebox(0,0)[lb]{\smash{\SetFigFont{12}{14.4}{rm}$p_3(X) \leftarrow p_1(X), p_1(X)$}}}
\put(5701,-3661){\makebox(0,0)[lb]{\smash{\SetFigFont{12}{14.4}{rm}ancestry with top clause}}}
\put(5176,-7111){\makebox(0,0)[lb]{\smash{\SetFigFont{12}{14.4}{rm}$\Box$}}}
\put(5551,-6661){\makebox(0,0)[lb]{\smash{\SetFigFont{12}{14.4}{rm}ancestry with top clause}}}
\put(6676,389){\makebox(0,0)[lb]{\smash{\SetFigFont{12}{14.4}{rm}$Res_{DB}: r(X) \leftarrow p_1(X), p_2(X)$}}}
\put(6751,-511){\makebox(0,0)[lb]{\smash{\SetFigFont{12}{14.4}{rm}$IC: p_2(X) \lor p_3(X) \leftarrow p_1(X)$}}}
\put(6751,-2311){\makebox(0,0)[lb]{\smash{\SetFigFont{12}{14.4}{rm}$Res_{DB}: r(X) \leftarrow p_3(X)$}}}
\put(6826,-5536){\makebox(0,0)[lb]{\smash{\SetFigFont{12}{14.4}{rm}$Res_{DB}: r(X) \leftarrow p_3(X)$}}}
\put(4876,-4111){\makebox(0,0)[lb]{\smash{\SetFigFont{12}{14.4}{rm}$ \leftarrow  p_1(X)$}}}
\put(4876,-2086){\makebox(0,0)[lb]{\smash{\SetFigFont{12}{14.4}{rm}$p_3(X) \leftarrow  p_1(X)$}}}
\put(4801,-3061){\makebox(0,0)[lb]{\smash{\SetFigFont{12}{14.4}{rm}$r(X) \leftarrow  p_1(X)$}}}
\put(6751,-4411){\makebox(0,0)[lb]{\smash{\SetFigFont{12}{14.4}{rm}$Query: p_1(k) \lor p_3(k) \leftarrow$}}}
\put(4801,-5161){\makebox(0,0)[lb]{\smash{\SetFigFont{12}{14.4}{rm}$p_3(k) \leftarrow$}}}
\put(4801,-6211){\makebox(0,0)[lb]{\smash{\SetFigFont{12}{14.4}{rm}$r(k) \leftarrow$}}}
\end{picture}
\caption{Use of Ancestry Resolution}
\end{figure}
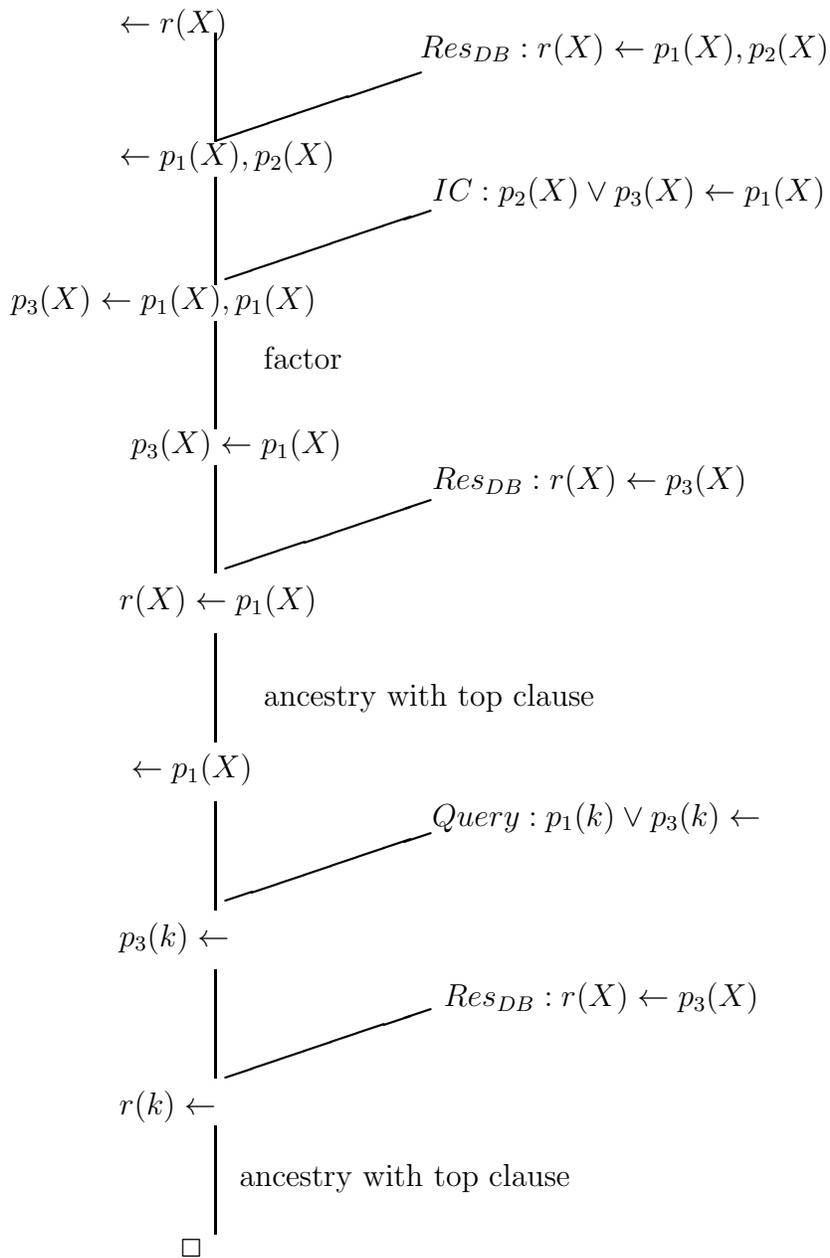


\section{Negation}\label{neg}
In this section we deal with stratified databases.  We allow default negation
in the {\bf IDB}, {\bf IC}, ${\bf Res_{DB}}$, and the query.  Otherwise our
assumptions on the database are the same as in Section~5.
Problems arise when we have just one resource 
rule that contains a conjunction of atoms, such as,
\begin{equation}
r(X) \leftarrow p_1(X), p_2(X)
\end{equation}
If we have the negation of $p_1(X)$ or $p_2(X)$ in the query with 
another atom, there is no way to resolve either of these atoms
with the $CCrr$ since the negated atoms do not appear
on the left hand side of any inverse rule.  To handle this, we represent a rule
for the negation of the resource as,
\begin{equation}
not~r(X) \leftarrow not~(p_1(X), p_2(X)).
\end{equation}
This would lead to two clauses
\begin{equation}
not~r(X) \leftarrow not~p_1(X),
\end{equation}
and
\begin{equation}
not~r(X) \leftarrow not~p_2(X)
\end{equation}
If we rename default negated atoms uniformly to new names with primes ('), we obtain:
\begin{equation}
r'(X) \leftarrow p'_1(X),
\end{equation}
and
\begin{equation}
r'(X) \leftarrow p'_2(X)
\end{equation}
Now, we have two rules that define $r'(X)$.  When we do the Clark completion
of this atom, we obtain a disjunctive inverse rule,
\begin{equation}
p'_1(X) \vee p'_2(X) \leftarrow r'(X)
\end{equation}
We must rename the negated atom in the query in the same manner.  Since we
have a disjunctive rule, as in Theorem~\ref{sound}, we may be in case (4)
and not be able to compute an answer.


We have to first deal with how to handle stratified databases.  The
underlying idea as developed by \cite{ABW88,vGe88} is that given a stratum,
and a negation of a predicate in the body of a rule in that stratum, the
predicate that is negated must have been defined in an earlier stratum, and hence,
the predicate may be calculated.  Thus its negation can be obtained.
An algorithm to make this explicit may be found in \cite{Ull88:principles} (Algorithm 3.6, Vol. 1).
We adapt this algorithm by compiling the {\bf IDB} predicates to rules
that involve {\bf EDB} predicates.
We give two versions of the compilation process.
In the first version, which requires a restriction on the types
of formulae allowed, we do a complete compilation, so that the
final rules involve {\bf EDB} predicates only.
In the second version we do not compile all {\bf IDB} predicates.

First, in order to do a complete compilation, we must restrict the type of 
formula that we allow for
defining {\bf IDB} and ${\bf Res_{DB}}$ predicates.
Namely, in addition to safety, we require that for all definitions the
set of variables in the body be identical to the set of variables in
the head. 
That is, there are no existential quantifiers in the right-hand side
of the formula.
We call such a formula {\em extra safe}.
By the definition of safe formula, no variable may appear in the head
that does not appear in the body.
Now we show why there should not be any variable in the body that is
not in the head.
Consider the case where the {\bf IDB} predicates $p$ and $t$ are
defined in terms of the {\bf EDB} predicates $h$, $k$, and $s$ as follows:
\begin{equation}
p(X,Y) \leftarrow h(X,Z), k(Z,Y)
\end{equation}
\begin{equation}
t(X,Y) \leftarrow s(X,Y), not~p(X,Y)
\end{equation}
Note that the definition of $p$ is safe but contains a variable $Z$
in the body that is not in the head, and hence is not extra safe.
In our compilation process, to be described below, we replace
$not~p(X,Y)$ by the negation of the body of the definition of $p$, so
we obtain the following two formulas as the compiled definition of $t$:
\begin{equation}
t(X,Y) \leftarrow s(X,Y), not~h(X,Z)
\end{equation}
\begin{equation}
t(X,Y) \leftarrow s(X,Y), not~k(Z,Y)
\end{equation}
obtaining two unsafe formulas as well as losing the connection between
$h$ and $k$ in $t$, namely, that $t$ is the join of $h$ and $k$.

Now we describe the compilation algorithm.
In a stratified database, each stratum is numbered in an increasing 
fashion.  Since there is no recursion in the {\bf IDB}, we can modify
the strata so that each stratum contains definitions for one predicate. 
For example, if $s$ and $t$ originally had definitions in the same
stratum and $s$ is the head of a rule that contains $t$ in the body of the rule,
we move the definitions for $s$ to the next higher stratum and adjust other stratum values
accordingly.  Call the predicate with definitions in stratum $i$, $p_{i}$.

\begin{description}
\item[Compilation Algorithm]\label{compile}
\item[Input:] {\bf EDB} (all predicates have stratum 0) and {\bf IDB} (predicates
with strata $1, \ldots n$).
\item[Output:] The compiled {\bf IDB} (each {\bf IDB} predicate defined in terms
of {\bf EDB} predicates).
\item[begin]
For stratum i = 1 To n Do (Note: {\bf EDB} predicates are not compiled)
\begin{enumerate}
\item If $p_i$ has multiple definitions, i.e., there are multiple rules with
$p_i$ as their head, replace these rules with a single definition
by taking the disjunction of the bodies of the multiple definitions.
\item Substitute for each predicate $p_j$, in the definition of $p_i$,
its compiled form.
\item Simplify the definition of $p_i$ by using De Morgan's rules and put
in disjunctive normal form.
\end{enumerate}
\item[end]
\end{description}

The following result shows that the Compilation Algorithm
does not change the extra safeness of predicates.
\begin{theorem}
\label{extrasafe}
Suppose that every predicate in {\bf IDB} is defined
by extra safe rules.
Then the compiled definitions are also extra safe rules.
\end{theorem}
\begin{proof}
We proceed by induction on the strata.
A predicate at stratum $0$ is extensional, and the statement is
vacuously true. Let $p_i$ be the IDB predicate at strata $i$.
By the inductive hypothesis, all predicates, $p_j$, $j < i$, 
in the body of a rule for $p_i$ have been compiled to extra safe
formulas.
Replacing the multiple extra safe rules in step 1 by a single rule
via a disjunction preserves the extra safe property.  Since
the compiled definition of each $p_j$ has exactly the same variables
as $p_j$, the substitution of step 2 also preserves the extra safe
property.  Finally, in step 3 using De Morgan's rules and converting
to disjunctive normal form preserves the extra safe property as well.
\end{proof} 

We now are able to handle stratified databases with extra safe rules
as given by the following theorem.

\begin{theorem}
\label{negation}
Let {\bf DB} be a stratified database where each {\bf IDB} rule is 
extra safe. 
Compile the {\bf IDB} predicates using the Compilation Algorithm to 
${\bf IDB^{C}}$.  
Apply ${\bf IDB^{C}}$ to the ${\bf Res_{DB}}$ rules to rewrite them in
terms of compiled {\bf IDB} predicates as  ${\bf Res^{C}_{DB}}$.
Rename every negated atom ($not~p$) in the query and in ${\bf Res^{C}_{DB}}$
to a new predicate ($p^{'}$) in a consistent manner and add 
the integrity constraints
$\leftarrow p, p^{'}$.  Then, the results of
Theorem 6.1 and Theorem~\ref{complete}, that deal with the leaves
of the proof tree, and the completeness test algorithm apply.
\end{theorem}

\begin{proof}
By Theorem~\ref{extrasafe} the Compilation Algorithm preserves extra 
safety for the predicates of  ${\bf IDB^{C}}$.  Hence, the variables
in ${\bf Res_{DB}}$ and ${\bf Res^{C}_{DB}}$ are the same.  By renaming
$not~p$ to $p^{'}$ we eliminate negation and the assumptions of 
Section 6 (see page 26) are satisfied.
The result follows because now, 
Theorem 6.1, and Theorem~\ref{complete} apply, and
the additional integrity constraints assure that it is not possible
to have both $p$ and $not~p$ at the same time.
\end{proof}

This theorem allows us to obtain proof trees and test for completeness.
In actually computing a folded query, we need to change back the primed
atoms $p^{'}$ to $not~p$.
The following example illustrates the theorem.
(We omit the transformation of $not~p$ to $p^{'}$.)

\begin{example}
Let the {\bf EDB} consist of: $e_{01}(X,Y)$, $e_{02}(X,Y,Z)$, 
$e_{03}(X,Y)$, and $e_{04}(X,Y)$.
Let the {\bf IDB} consist of the following clauses:
\begin{equation}
e_{11}(X,Y,Z) \leftarrow e_{01}(X,Y), e_{02}(X,Y,Z)
\end{equation}
\begin{equation}
e_{11}(X,Y,Z) \leftarrow e_{02}(X,Y,Z), not~e_{03}(X,Y)
\end{equation}
\begin{equation}
e_{12}(X,Y) \leftarrow e_{04}(X,Y), not~e_{01}(X,Y)
\end{equation}
\begin{equation}
e_{21}(X,Y) \leftarrow e_{03}(X,Y), not~e_{12}(X,Y)
\end{equation}
Now, let ${\bf Res_{DB}}$ consist of:
\begin{equation}
r(X,Y,Z) \leftarrow e_{11}(X,Y,Z), not~e_{12}(X,Y)
\end{equation}
The above database is not compiled.  Using the algorithm given above,
the following database is found, where all rules are written in terms of
{\bf EDB} predicates.  The compiled definitions, ${\bf IDB^c}$ become: 
\begin{equation}
e_{11}(X,Y,Z) \leftarrow e_{02}(X,Y,Z), (e_{01}(X,Y) \vee not~e_{03}(X,Y))
\end{equation}
\begin{equation}
e_{21}(X,Y) \leftarrow e_{03}(X,Y), (not~e_{04}(X,Y) \vee e_{01}(X,Y))
\end{equation}
The compiled resource, ${\bf Res^{C}_{DB}}$ becomes:
\begin{equation}
r(X,Y,Z) \leftarrow e_{02}(X,Y,Z), (e_{01}(X,Y) \vee not~e_{03}(X,Y)), 
(e_{01}(X,Y), \vee not~e_{04}(X,Y))
\end{equation}
The Clark Completion resource rules become:
\begin{equation}
CCrr1: e_{02}(X,Y,Z) \leftarrow r(X,Y,Z)
\end{equation}
\begin{equation}
CCrr2: e_{01}(X,Y) \vee not~e_{03}(X,Y) \leftarrow r(X,Y,Z)
\end{equation}
\begin{equation}
CCrr3: e_{01}(X,Y) \vee not~e_{04}(X,Y,Z) \leftarrow r(X,Y,Z)
\end{equation}
Let the negation of the query be: $Q: \leftarrow e_{21}(X,Y)$
 
In 
Figure 11, 
we show that using the query and
the Clark Completion resource rules yields a partial folding
on the leftmost branch:
\begin{equation}
\leftarrow r(X,Y,Z), e_{03}(X,Y)
\end{equation}

\begin{figure}
\label{stratified-example}
\setlength{\unitlength}{0.00083300in}%
\begingroup\makeatletter\ifx\SetFigFont\undefined
\def\x#1#2#3#4#5#6#7\relax{\def\x{#1#2#3#4#5#6}}%
\expandafter\x\fmtname xxxxxx\relax \def\y{splain}%
\ifx\x\y   
\gdef\SetFigFont#1#2#3{%
  \ifnum #1<17\tiny\else \ifnum #1<20\small\else
  \ifnum #1<24\normalsize\else \ifnum #1<29\large\else
  \ifnum #1<34\Large\else \ifnum #1<41\LARGE\else
     \huge\fi\fi\fi\fi\fi\fi
  \csname #3\endcsname}%
\else
\gdef\SetFigFont#1#2#3{\begingroup
  \count@#1\relax \ifnum 25<\count@\count@25\fi
  \def\x{\endgroup\@setsize\SetFigFont{#2pt}}%
  \expandafter\x
    \csname \romannumeral\the\count@ pt\expandafter\endcsname
    \csname @\romannumeral\the\count@ pt\endcsname
  \csname #3\endcsname}%
\fi
\fi\endgroup
\begin{picture}(9837,6942)(1,-6247)
\thicklines
\put(2401,-736){\line(-5,-3){1213.235}}
\put(9826,-3511){\makebox(6.6667,10.0000){\SetFigFont{10}{12}{rm}.}}
\put(1276,-436){\line( 5, 1){2250}}
\put(676,389){\line( 0,-1){825}}
\put(676,-1786){\line( 0,-1){825}}
\put(676,-2911){\line( 0,-1){825}}
\put(2776,-736){\line( 0, 1){  0}}
\put(2926,-811){\line( 4,-1){2982.353}}
\put(751,-2611){\line( 4, 1){1658.823}}
\put(2926,-1786){\makebox(6.6667,10.0000){\SetFigFont{10}{12}{rm}.}}
\put(3601,-2161){\line(-1,-5){750}}
\put(2926,-5911){\line( 5, 6){3405.738}}
\put(1501,-3736){\line( 5, 2){4461.207}}
\put(5626,-4636){\line( 0,-1){600}}
\put(6849,-1868){\line(-1,-4){851.471}}
\put(676,-4036){\line( 0,-1){825}}
\put(  1,539){\makebox(0,0)[lb]{\smash{\SetFigFont{12}{14.4}{rm}$Query: \leftarrow e_{21}(X,Y)$}}}
\put(2176,164){\makebox(0,0)[lb]{\smash{\SetFigFont{12}{14.4}{rm}$IDB^C: e_{21}(X,Y) \leftarrow e_{03}(X,Y) (not~e_{04}(X,Y) \vee e_{01}(X,Y))$}}}
\put( 76,-2836){\makebox(0,0)[lb]{\smash{\SetFigFont{12}{14.4}{rm}$e_{01}(X,Y) \leftarrow r(X,Y,Z), e_{03}(X,Y)$}}}
\put(5176,-1786){\makebox(0,0)[lb]{\smash{\SetFigFont{12}{14.4}{rm}$\leftarrow e_{03}(X,Y), e_{01}(X,Y)$}}}
\put(3976,-4561){\makebox(0,0)[lb]{\smash{\SetFigFont{12}{14.4}{rm}$CCrr2: e_{01}(X,Y) \vee not~e_{03}(X,Y) \leftarrow r(X,Y,Z)$}}}
\put(4501,-5461){\makebox(0,0)[lb]{\smash{\SetFigFont{12}{14.4}{rm}$not~e_{03}(X,Y) \leftarrow r(X,Y,Z), e_{03}(X,Y)$}}}
\put(  1,-1711){\makebox(0,0)[lb]{\smash{\SetFigFont{12}{14.4}{rm}$\leftarrow e_{03}(X,Y), not~e_{04}(X,Y)$}}}
\put( 76,-736){\makebox(0,0)[lb]{\smash{\SetFigFont{12}{14.4}{rm}$\leftarrow e_{03}(X,Y) (not~e_{04}(X,Y) \vee e_{01}(X,Y))$}}}
\put(826,-2086){\makebox(0,0)[lb]{\smash{\SetFigFont{12}{14.4}{rm}$CCrr3: e_{01}(X,Y) \vee not~e_{04}(X,Y) \leftarrow r(X,Y,Z)$}}}
\put(1801,-6211){\makebox(0,0)[lb]{\smash{\SetFigFont{12}{14.4}{rm}$not~e_{04}(X,Y) \leftarrow r(X,Y,Z), e_{03}(X,Y)$}}}
\put( 76,-3961){\makebox(0,0)[lb]{\smash{\SetFigFont{12}{14.4}{rm}$\leftarrow r(X,Y,Z), e_{03}(X,Y),e_{03}(X,Y)$}}}
\put( 76,-5086){\makebox(0,0)[lb]{\smash{\SetFigFont{12}{14.4}{rm}$\leftarrow r(X,Y,Z), e_{03}(X,Y)$}}}
\put(826,-4486){\makebox(0,0)[lb]{\smash{\SetFigFont{12}{14.4}{rm}factor}}}
\end{picture}
\caption{Stratification Example}
\end{figure}

\end{example}

We now consider the second case where the rules are safe but not
necessarily extra safe.
In this case modify the Compilation Algorithm in step 2 so that
predicates with a definition that is not extra safe are not
compiled.
In this case the end result of the compilation may contain
{\bf IDB} predicates in some definitions.
We use the same notation for the compiled versions, i.e.
${\bf IDB^C}$ and ${\bf Res_{DB}^C}$ as before.
Theorem~\ref{negation} extends to this case as follows.
\begin{theorem}
Let {\bf DB} be a stratified database.
Compile the {\bf IDB} predicates 
using the Compilation Algorithm modified as explained above to
${\bf IDB^C}$.
Proceed as in the statement of Theorem~\ref{negation},
compiling ${\bf Res_{DB}}$ to ${\bf Res_{DB}^C}$ and renaming
the negated predicates.
Then, the results of
Theorem~5.1 and Theorem~\ref{complete} apply.
\end{theorem}
\begin{proof}
Similar to the proof of Theorem~\ref{negation} except that the 
modified Compilation Algorithm is used and so the compilation
process stops earlier for certain predicates.
\end{proof}
The following example illustrates the Theorem.
(Again we omit the transformation of $not~p$ to $p'$.)
\begin{example}
Let the {\bf EDB} consist of: $h(X,Y)$, $k(X,Y)$, 
$s(X,Y)$, and $\ell(X,Y,Z)$.
Let the {\bf IDB} consist of the following clauses:
\begin{equation}
p(X,Y) \leftarrow h(X,Z), k(Z,Y)
\end{equation}
\begin{equation}
t(X,Y) \leftarrow s(X,Y), not~p(X,Y)
\end{equation}
Now, let ${\bf Res_{DB}}$ consist of:
\begin{equation}
r(X,Y,Z) \leftarrow t(X,Y), \ell(Y,Z,U)
\end{equation}
Note that the definition for $p$ is not extra safe.
So in the definition of $t(X,Y)$, $not~p(X,Y)$ is not changed.
Thus, ${\bf IDB^C}$ = {\bf IDB}
and the compiled resource, ${\bf Res^{c}_{DB}}$ becomes:
\begin{equation} 
r(X,Y,Z) \leftarrow s(X,Y), not~p(X,Y), \ell(Y,Z,U)
\end{equation}
The Clark Completion resource rules become:
\begin{equation}
CCrr1: s(X,Y) \leftarrow r(X,Y,Z)
\end{equation}
\begin{equation}
CCrr2: not~p(X,Y) \leftarrow r(X,Y,Z)
\end{equation}
\begin{equation}
CCrr3: \ell(Y,Z,f) \leftarrow r(X,Y,Z)
\end{equation}
Let the negation of the query be: $Q: \leftarrow s(X,Y), k(Y,Z)$
\end{example}
 
In Figure 12
we show that using the query and 
the Clark Completion resource rules yields a partial folding:
\begin{equation}
\leftarrow r(X,Y,Z), k(Y,Z)
\end{equation}

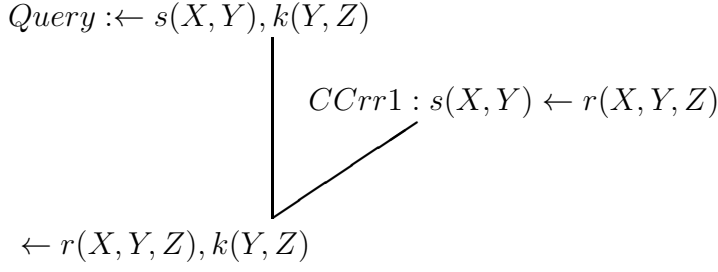
\begin{figure}
\label{safe}
\setlength{\unitlength}{0.00083300in}%
\begingroup\makeatletter\ifx\SetFigFont\undefined
\def\x#1#2#3#4#5#6#7\relax{\def\x{#1#2#3#4#5#6}}%
\expandafter\x\fmtname xxxxxx\relax \def\y{splain}%
\ifx\x\y   
\gdef\SetFigFont#1#2#3{%
  \ifnum #1<17\tiny\else \ifnum #1<20\small\else
  \ifnum #1<24\normalsize\else \ifnum #1<29\large\else
  \ifnum #1<34\Large\else \ifnum #1<41\LARGE\else
     \huge\fi\fi\fi\fi\fi\fi
  \csname #3\endcsname}%
\else
\gdef\SetFigFont#1#2#3{\begingroup
  \count@#1\relax \ifnum 25<\count@\count@25\fi
  \def\x{\endgroup\@setsize\SetFigFont{#2pt}}%
  \expandafter\x
    \csname \romannumeral\the\count@ pt\expandafter\endcsname
    \csname @\romannumeral\the\count@ pt\endcsname
  \csname #3\endcsname}%
\fi
\fi\endgroup
\begin{picture}(2562,1617)(3526,-922)
\thicklines
\put(5176,-661){\line( 3, 2){900}}
\put(5176,464){\line( 0,-1){1125}}
\put(3526,539){\makebox(0,0)[lb]{\smash{\SetFigFont{12}{14.4}{rm}$Query: \leftarrow s(X,Y), k(Y,Z)$}}}
\put(5401, 14){\makebox(0,0)[lb]{\smash{\SetFigFont{12}{14.4}{rm}$CCrr1: s(X,Y) \leftarrow r(X,Y,Z)$}}}
\put(3601,-886){\makebox(0,0)[lb]{\smash{\SetFigFont{12}{14.4}{rm}$\leftarrow r(X,Y,Z), k(Y,Z)$}}}
\end{picture}
\caption{Safe Negation Example}
\end{figure}

We end this section by showing that any folded query resulting from
Theorems 6.2 or 6.3 is safe.
We start by proving a general result.
\newtheorem{prop}{Proposition}[section]
\begin{prop}
The resolution of two safe formulas (in Datalog with negation and 
disjunction) is a safe formula.
\end{prop}
\begin{proof}
Let the two safe formulas have the form:\\
$A_1, \ldots, A_k \leftarrow A_{k+1}, \ldots, A_m$ and\\
$B_1, \ldots, B_{\ell} \leftarrow B_{\ell + 1}, \ldots, B_n$,\\
where each $A_i$, $B_j$, $1 \leq i \leq m$, $1 \leq j \leq n$ is an atom.
We may assume without loss of generality that $A_1$ is resolved with $B_n$
to yield\\
$A_2, \ldots, A_k, B_1, \ldots, B_{\ell} \leftarrow
A_{k+1}, \ldots, A_m, B_{\ell + 1}, \ldots, B_{n-1}$.\\
Actually, the resolution involves a substitution $\theta$, so here each
$A_i$, $B_j$, $2 \leq i \leq m$, $1 \leq j \leq n-1$ is really
$A_i \theta$, $B_j \theta$, but this can be ignored for the purpose of
this proof because any changed variable in the head of a clause must be
changed the same way in the body.

We need to show that every variable in the resulting formula is limited.
If $X$ is a variable in any $A_i$, $2 \leq i \leq m$, it must already
be limited in $A_{k+1}, \ldots, A_m$ because the $A$ formula is safe.
If $X$ is a variable in any $B_j$, $1 \leq j \leq n-1$, that did not appear
in $B_n$, it must already be limited in $B_{\ell+1}, \ldots, B_{n-1}$
because the $B$ formula is safe.
Finally, if $X$ is a variable in some $B_j$, $1 \leq j \leq n-1$, 
that was limited in the $B$ formula in $B_n$, by the resolution $X$
must now be limited in $A_{k+1}, \ldots, A_m$.
\end{proof}
\newtheorem{cor}{Corollary}
\begin{cor}
Every folded query resulting from Theorem 6.3 is safe.
\end{cor}
\begin{proof}
The query is safe and so are the rules in {\bf IDB} and {\bf IC}.
Consider the way the Clark Completion resource rules are obtained.
In each such rule the body contains only the resource predicate and
every additional variable in the body of an original resource rule becomes a 
function symbol.
Since these function symbols cannot be iterated, they can be treated
as constants.
Hence the Clark Completion resource rules are also safe and the
result follows from the Proposition.
\end{proof}

\section{Recursion}\label{recur}
Query folding becomes problematic in the presence of recursion,
since one does not know when to terminate recursion in a top-down
approach.  However, if it is known that the recursion is bounded 
\cite{Mink82A,Naug87}, that is, the recursion is known to terminate
after a number of stages using only intensional rules, then one can use the methods we describe
in the previous sections to handle this case.  In this section we assume that
recursion is not bounded.  

Unlike the previous sections, the computations we present here follow
the bottom-up approach rather than the top-down method we used earlier.
The reason is that in the presence of recursion it is difficult to
tell when all the solutions have been obtained in the infinite
proof tree.  In this case, we are not doing query folding, but we
are doing query answering.  That is, we do not find a rewriting
of the query in terms of the resources.
Actually, there are strong connections between the top-down and
bottom-up approaches: Bry \cite{Bry90} describes a combined top-down,
bottom-up interpreter that incorporates the magic set technique used
for recursion.

We start with the case where the recursion occurs only in the query,
so the {\bf IDB}, {\bf IC}, and ${\bf Res_{DB}}$ contain no recursion.
This is equivalent to the case where recursion appears in the {\bf IDB},
a query is asked in terms of the {\bf IDB} and the ${\bf Res_{DB}}$
definitions include only {\bf EDB} predicates (and {\bf IDB} predicates
defined in such a way that they can be compiled to {\bf EDB} predicates
without recursion).

Within the case where recursion only occurs in the query we start with
the subcase where everything is positive, that is, there is no negation.
The database restrictions are as in Section 4.1 except that the query
is recursive.
This case was solved in \cite{DGL00:integration}.
We start by considering their Example 3.1.\\
\begin{example}
{\bf EDB}: $edge(X,Y)$
{\bf IDB}: $\emptyset$\\
{\bf IC}: $\emptyset$\\
${\bf Res_{DB}}$:  $r(X,Y) \leftarrow edge(X,Z), edge(Z,Y)$\\
{\bf Query}: $q(X,Y) \leftarrow edge(X,Y)$\\
\hspace*{.55in} $q(X,Y) \leftarrow edge(X,Z), q(Z,Y)$\\
\end{example}

In this example the recursive query determines the transitive closure
of the relation $edge$, while the resource predicate stores endpoints
of paths of length 2.
The Clark Completion resource rules here are\\\\
$CCrr1: edge(X,f(X,Y)) \leftarrow r(X,Y)$.\\\\
$CCrr2: edge(f(X,Y),Y) \leftarrow r(X,Y)$.\\\\
In this type of situation the solution is  a two-step process.
In the first step the extensional 
predicates, {\em edge} in this case, 
are evaluated from the resource predicates
using the Clark Completion resource rules, and
in the second step a bottom-up Datalog evaluation is done for the
recursive query from the extensional predicates.
It is shown there why this process always terminates in a finite
amount of time.  Specifically, they note that the key observation is
that function symbols are only introduced in inverse rules.  Because
inverse rules are not recursive, no terms with nested function symbols
can be generated.
As we mentioned earlier the problem with the top-down approach is
that it is difficult to find out when to stop processing.
We draw one branch of the top-down tree in Figure 13. 

\begin{figure}
\label{onebranch}
\setlength{\unitlength}{0.00083300in}%
\begingroup\makeatletter\ifx\SetFigFont\undefined
\def\x#1#2#3#4#5#6#7\relax{\def\x{#1#2#3#4#5#6}}%
\expandafter\x\fmtname xxxxxx\relax \def\y{splain}%
\ifx\x\y   
\gdef\SetFigFont#1#2#3{%
  \ifnum #1<17\tiny\else \ifnum #1<20\small\else
  \ifnum #1<24\normalsize\else \ifnum #1<29\large\else
  \ifnum #1<34\Large\else \ifnum #1<41\LARGE\else
     \huge\fi\fi\fi\fi\fi\fi
  \csname #3\endcsname}%
\else
\gdef\SetFigFont#1#2#3{\begingroup
  \count@#1\relax \ifnum 25<\count@\count@25\fi
  \def\x{\endgroup\@setsize\SetFigFont{#2pt}}%
  \expandafter\x
    \csname \romannumeral\the\count@ pt\expandafter\endcsname
    \csname @\romannumeral\the\count@ pt\endcsname
  \csname #3\endcsname}%
\fi
\fi\endgroup
\begin{picture}(3162,5442)(3001,-4972)
\thicklines
\put(4201,314){\line( 0,-1){450}}
\put(4201,-361){\line( 0,-1){975}}
\put(4201,-1336){\line( 3, 1){1935}}
\put(4201,-1636){\line( 0,-1){975}}
\put(4201,-2590){\line( 3, 1){1935}}
\put(4201,-2986){\line( 0,-1){975}}
\put(4201,-3912){\line( 3, 1){1935}}
\put(4201,-4261){\line( 0,-1){450}}
\put(3001,314){\makebox(0,0)[lb]{\smash{\SetFigFont{12}{14.4}{rm}Query: $\leftarrow q(X,Y)$}}}
\put(4501, 14){\makebox(0,0)[lb]{\smash{\SetFigFont{12}{14.4}{rm}$q(X,Y) \leftarrow edge(X,Z), q(Z,Y)$}}}
\put(3601,-361){\makebox(0,0)[lb]{\smash{\SetFigFont{12}{14.4}{rm}$\leftarrow edge(X,Z), q(Z,Y)$}}}
\put(4576,-661){\makebox(0,0)[lb]{\smash{\SetFigFont{12}{14.4}{rm}CCrr1: $edge(X^{'},f(X^{'},Y^{'})) \leftarrow r(X^{'},Y^{'})$}}}
\put(4576,-1936){\makebox(0,0)[lb]{\smash{\SetFigFont{12}{14.4}{rm}$q(X^{''},Y^{''}) \leftarrow edge(X^{''},Y^{''})$}}}
\put(5551,-2461){\makebox(0,0)[lb]{\smash{\SetFigFont{12}{14.4}{rm}$\{X^{''}/f(X,Y^{'}), Y^{''}/Y\}$}}}
\put(4576,-3211){\makebox(0,0)[lb]{\smash{\SetFigFont{12}{14.4}{rm}CCrr2: $edge(f(X^{''},Y^{''}), Y^{''}) \leftarrow r(X^{''},Y^{''})$}}}
\put(5551,-3736){\makebox(0,0)[lb]{\smash{\SetFigFont{12}{14.4}{rm}$\{X^{''}/X, Y^{''}/Y, Y^{'}/Y\}$}}}
\put(4576,-4561){\makebox(0,0)[lb]{\smash{\SetFigFont{12}{14.4}{rm}factor}}}
\put(3601,-4186){\makebox(0,0)[lb]{\smash{\SetFigFont{12}{14.4}{rm}$\leftarrow r(X,Y), r(X,Y)$}}}
\put(3601,-1561){\makebox(0,0)[lb]{\smash{\SetFigFont{12}{14.4}{rm}$\leftarrow r(X,Y^{'}), q(f(X,Y^{'}),Y)$}}}
\put(5476,-1111){\makebox(0,0)[lb]{\smash{\SetFigFont{12}{14.4}{rm}$\{Z/f(X,Y^{'}), X^{'}/X\}$}}}
\put(3601,-2836){\makebox(0,0)[lb]{\smash{\SetFigFont{12}{14.4}{rm}$\leftarrow r(X,Y^{'}), edge(f(X,Y^{'}),Y)$}}}
\put(3601,-4936){\makebox(0,0)[lb]{\smash{\SetFigFont{12}{14.4}{rm}$\leftarrow r(X,Y)$}}}
\end{picture}
\caption{One branch of the recursive proof tree}
\end{figure}
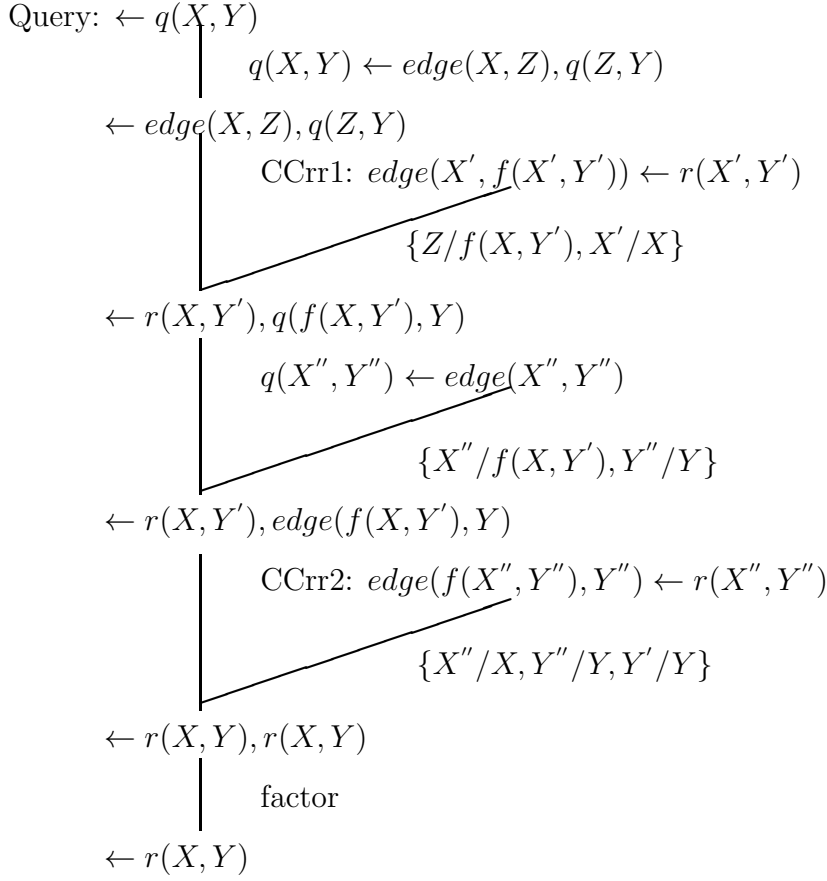

The second subcase is where the recursive query is stratified, so negation
is allowed.
A similar two-step process works here also.
The first step is the same as before.
However, in the second step the query is computed by computing the
stratified database \cite{Ull88:principles}.

The final subcase we consider is where the resources are defined by
multiple definitions (but still without recursion) as in Section 6.
Recall that now the Clark Completion resource rules will contain
disjunctions in the head.
We illustrate the idea by reconsidering the simple example from the
beginning of Section 6 where one Clark completion resource rule
(written as a single rule) is:
\begin{equation}
(p_1(X) \wedge p_2(X)) \vee p_3(X) \leftarrow r(X)
\end{equation}
Suppose we have $r(a)$.
Then we separately do three subcomputations:
\begin{itemize}
\item generate $p_1(a)$ and $p_2(a)$,
\item generate $p_3(a)$,
\item generate $p_1(a)$, $p_2(a)$, and $p_3(a)$.
\end{itemize}
In Step 1 apply this process to all the Clark Completion resource
rules to obtain all the subcomputations.
For every subcomputation apply Step 2, the bottom-up Datalog
evaluation of the query.
Place a tuple in the answer only if it is an answer in every
subcomputation.
This is essentially the minimal models approach for disjunctive
logic programming \cite{LMR92:book}.

We end our exploration of recursion by considering the case where
recursion is allowed in the ${\bf Res_{DB}}$.
In general, there is no known method to handle this.  In the following,
we discuss a specific form of recursion for which we obtain sound answers.
The problem is how to get the Clark Completion resource rules in a 
usable form to give us definitions of extensional predicates in
terms of resource predicates.
Going back to the previous $edge$ example, suppose that instead of
the resource predicate storing endpoints of paths of length 2, the
resource predicate stores the endpoints of all paths.
This would be written as\\
${\bf Res_{DB}}$: $r(X,Y) \leftarrow edge(X,Y)$\\
\hspace*{.5in} $r(X,Y) \leftarrow edge(X,Z), r(Z,Y)$\\
Now we write a modified Clark Completion resource rule as:
\begin{equation}
MCCrr: edge(X,Y) \leftarrow r(X,Y), not \exists Z (r(X,Z), r(Z,Y))
\end{equation}
obtaining all those paths in $r$ that cannot be broken up into two
paths.
The two steps of the computation process are as before: evaluate
$edge$ first from $r$ (using the modified Clark Completion resource
rule) and then evaluate $q$ from $edge$ whether or not $q$ is
recursive.
The general result is as follows:
\newtheorem{proposition}{Proposition}[section]
\begin{proposition}
Suppose that a ${\bf Res_{DB}}$ predicate $r$ is recursive 
and has the form\\
$r(\bar{X}) \leftarrow e(\bar{X})$\\
$r(\bar{X}) \leftarrow t(\bar{Y})$\\
where $t$ is a conjunction that may include $e$ and $r$ 
and $\bar{X} \subseteq \bar{Y}$\\
Then the modified Clark Completion resource rule is:\\
$MCCrr: e(\bar{X}) \leftarrow r(\bar{X}), not \exists \bar{Z} t(\bar{Y})$\\
where $\bar{Z} = \bar{Y} - \bar{X}$.\\
$MCCrr$ can be used to evaluate $e$ from $r$ in a sound manner.
\end{proposition}
\begin{proof}
We proceed by contraposition. 
Suppose that for some tuple $\bar{a}$ 
$e(\bar{a})$ is false and $r(\bar{a})$ is true.
This means that $r(\bar{a})$ must have been obtained by the second, 
(recursive) definition.
But then $\exists \bar{Z} t(\bar{Y})$ must be true.
\end{proof}

\section{Comparison}\label{comp}
In this section we discuss the contributions made in this paper and compare
the work with other efforts. 

Qian (\cite{Q96}) was the first to consider the problem of folding. 
In her paper she introduced the concept of inverse rules to permit one
to use resources to compute answers to queries.  Inverse rules basically
state that the only way in which the information in the resource can be
computed is through the resource.  As noted in  this paper, the concept 
of inverse rules was 
introduced first in the context of logic programming by Clark \cite{Q96}, 
and is the basis of the {\em closed world assumption}.  It represents
an if-and-only-if condition for rules.
Qian showed that if, for each resource predicate, defined by a conjunctive rule,
there is at most one such rule, that it is possible to compute the answer
to queries using resources in many instances.  We have extended Qian's result
slightly to include databases that contain arbitrary integrity constraints.
Duschka and his associates \cite{Duschka97:AAAI,DG97:ans-queries-views,Duschka97} show 
how to extend the work to classes of integrity constraints.  
\cite{DG97:ans-queries-views,Duschka97} were the first to extend
the work to handle general recursive queries.

Dawson, Gryz and Qian \cite{DGQ96} show how to compute answers in query
folding when there are functional dependencies.  \cite{DL97,Duschka97} 
introduced the new class of {\em recursive} query plans for information
gathering.  Instead of plans being only sets of conjunctive queries, they
can now be recursive sets of function-free Horn clauses.  Using recursive
plans, they settle two open problems.  First, they describe an algorithm
for finding the maximally contained rewriting in the presence of functional
dependencies.  Second, they describe an algorithm for finding the
maximally-contained rewriting in the presence of binding-pattern restrictions,
which was not possible without recursive plans.  
We have shown how integrity constraints containing equality, such as 
functional dependencies, generate a possibly infinite set of folded queries.

Duschka and Genesereth \cite{DG98:disj,Duschka97} developed the first 
algorithm to solve the problem of answering queries using views when view 
definitions are allowed to contain disjunction.  They use the Clark completion
to obtain the inverse rules.  Disjunctive definitions for rules implies that
$Datalog$ is not sufficiently powerful to handle such situations and it is
necessary to use $Datalog^{\vee}$.  Their focus is on maximal query 
containment. 
They show a duality in between a query plan being maximally contained in a 
query and this plan computing exactly the certain answers. They show that the 
disjunctive plan can be evaluated in co-NP time.
Afrati et al.  \cite{AGK99:disjviews} also treats the problem of disjunctive 
materialized views. 
The relationship with our approach is that we show how, 
using theorem proving concepts we
can handle such theories, including integrity constraints.  We also show how
one can determine if the query plan that has been developed is complete, that
is, if all answers to the original query have been obtained.  We do this in 
the context of theorem proving.

We know of no work that covers negation in the folding problem.  We show
how to handle stratified negation in views which may be defined by 
disjunctive rules, and may contain integrity constraints.  Duschka 
\cite{Duschka97} discusses complexity results with respect to 
negation in his thesis.

With respect to recursion, as noted above, \cite{DG97:ans-queries-views} handle recursion.
We show how to handle recursion in a similar manner to their work, and extend
the results to stratified databases.  In general, there is no known method
to handle recursion in resource rules.  We suggest one limited case where
recursion appears in resource rules and sound answers may be found.  


\section{Summary}\label{summ}
We have shown that a logic-based approach using resolution unifies 
techniques used earlier for the aspect of data integration also known as
query folding.
We considered a deductive database where a query written on the database
needs to be rewritten in terms of given resources.
We showed how to handle integrity constraints and the case where a
resource has multiple definitions.
We also showed when the folded query yields all or only some of the
answers of the original query.
We extended our results to some cases involving negation and recursion.

\section*{Acknowledgements}\label{ack}
We would like to express our appreciation to Parke Godfrey 
and Jarek Gryz for many discussions we have had with them
and for their suggestions.  We also wish to thank Alon Levy 
who made several valuable suggestions which helped to
improve the paper.  We are also grateful to the referees for
their many helpful comments and suggestions.

\bibliographystyle{alpha}
\bibliography{/jacksun/minker/BIB/biblio,%
/jacksun/minker/BIB/database,%
/jacksun/minker/BIB/extendedlp,%
/jacksun/minker/BIB/main,%
/jacksun/minker/BIB/nicolas,%
/jacksun/minker/BIB/ns,%
/jacksun/minker/BIB/new,%
/jacksun/minker/BIB/newns,%
/jacksun/minker/BIB/old.biblio,%
/jacksun/minker/BIB/parke,%
/jacksun/minker/BIB/book,%
/jacksun/minker/DOCUMENTS/PAPERS/1991/nonmon}

\end{document}